\documentclass[12pt]{iopart}
\eqnobysec
\usepackage{iopams} 
\usepackage{amsthm} 

\newcommand{\Sw}{{\cal S}(\mathbb R \frac{\ }{\ } \{ a,b \} )}
\newcommand{\Swt}{{\cal S}^{\times}(\mathbb R \frac{\ }{\ } \{ a,b \} )}
\newcommand{\Swp}{{\cal S}^{\prime}(\mathbb R \frac{\ }{\ } \{ a,b \} )}
\newcommand{\rhsSwt}{\Sw \subset L^2(\mathbb R, \rmd x) \subset \Swt}
\newcommand{\rhsSwp}{\Sw \subset L^2(\mathbb R, \rmd x) \subset \Swp}
\newcommand{\Czi}{C_0^{\infty}(\mathbb R \frac{\ }{\ } \{ a,b \} )}
\newcommand{\Swhpm}{\widehat{{\cal S}}_{\pm}(\mathbb R \frac{\ }{\ }\{ a,b \})}
\newcommand{\Swhpmt}{\widehat{{\cal S}}_{\pm}^{\times}(\mathbb R \frac{\ }{\ }\{ a,b \})}
\newcommand{\Swhpml}{\widehat{{\cal S}}_{\pm;l}(\mathbb R \frac{\ }{\ }\{ a,b \})}
\newcommand{\Swhpmr}{\widehat{{\cal S}}_{\pm;r}(\mathbb R \frac{\ }{\ }\{ a,b \})}
\newcommand{\Swhpmlt}{\widehat{{\cal S}}_{\pm;l}^{\times}(\mathbb R \frac{\ }{\ }\{ a,b \})}
\newcommand{\Swhpmrt}{\widehat{{\cal S}}_{\pm;r}^{\times}(\mathbb R \frac{\ }{\ }\{ a,b \})}
\newcommand{\Swhpmp}{\widehat{{\cal S}}_{\pm}^{\prime}(\mathbb R \frac{\ }{\ }\{ a,b \})}
\newcommand{\Swhpmlp}{\widehat{{\cal S}}_{\pm;l}^{\prime}(\mathbb R \frac{\ }{\ }\{ a,b \})}
\newcommand{\Swhpmrp}{\widehat{{\cal S}}_{\pm;r}^{\prime}(\mathbb R \frac{\ }{\ }\{ a,b \})}
\newcommand{\Swh}{ \widehat{{\cal S}}(\mathbb R \frac{\ }{\ } \{ a,b \} )}
\newcommand{\Swht}{ \widehat{{\cal S}}^{\times}(\mathbb R \frac{\ }{\ } \{ a,b \} )}
\newcommand{\Swhp}{\widehat{{\cal S}}^{\prime}(\mathbb R \frac{\ }{\ } \{ a,b \} )}
\newcommand{\Swhhpm}{\widehat{\widehat{{\cal S}}\,}\! _{\pm}(\mathbb R \frac{\ }{\ } \{ a,b \} )}
\newcommand{\Swhhpmt}{\widehat{\widehat{{\cal S}}\,}\! _{\pm} \, \hskip-.32cm^{\times}(\mathbb R \frac{\ }{\ } \{ a,b \} )}
\newcommand{\Swhhpmp}{\widehat{\widehat{{\cal S}}\,}\! _{\pm} \,\hskip-.22cm ^{\prime}(\mathbb R \frac{\ }{\ } \{ a,b \} )}

\begin{document}

\def\llra{\relbar\joinrel\longrightarrow}              
\def\mapright#1{\smash{\mathop{\llra}\limits_{#1}}}    
\def\mapup#1{\smash{\mathop{\llra}\limits^{#1}}}     
\def\mapupdown#1#2{\smash{\mathop{\llra}\limits^{#1}_{#2}}} 

\title[The RHS of the 1D rectangular barrier potential]{The rigged Hilbert 
space of the algebra of the one-dimensional rectangular barrier potential}

\author{Rafael de la Madrid}

\address{
Departamento de F\'\i sica Te\'orica, Facultad de Ciencias,
Universidad del Pa\'\i s Vasco \\ E-48080 Bilbao, Spain \\
E-mail: {\texttt{wtbdemor@lg.ehu.es}} \\
URL: {\texttt{http://www.ehu.es/$\sim$wtbdemor}}}

\begin{abstract}
The rigged Hilbert space of the algebra of the one-dimensional rectangular 
barrier potential is constructed. The one-dimensional rectangular
potential provides another opportunity to show that the rigged Hilbert space 
fully accounts for Dirac's bra-ket formalism. The analogy between Dirac's 
formalism and Fourier methods is pointed out.
\end{abstract}



\maketitle

\section{Introduction}

One-dimensional (1D) models play a paramount role in Quantum Mechanics, because
they enable us to understand a number of properties that also appear in more 
realistic situations. The simplicity of 1D models facilitates testing 
new hypothesis, approximation methods and theories without unnecessary and 
costly complications. In many cases, after proper calculations, it is 
possible to reduce an intricate problem to a Schr\"odinger equation in one 
dimension. For example: three-dimensional (3D) spherically symmetric 
Schr\"odinger equations can be reduced to 1D radial equations; time quantities 
such as tunneling or arrival times have in many cases been studied in 
1D models~\cite{LEON,MUGA}; electrons in strong magnetic fields
can be described by 1D potentials~\cite{BRUMMELHUIS}; some surface 
phenomena are described by 1D models~\cite{BY}; the application of the 
effective mass approximation to layered semiconductor structures leads to 
effective 1D systems~\cite{BASTARD}; the conductance of some semiconductor
nanostructures can be obtained by solving 1D Schr\"odinger 
equations~\cite{WULF}. One-dimensional potentials 
have even practical interest, since advances in the microfabrication of 
semiconductors have allowed to design and control essentially 1D 
potentials~\cite{SAKAKI,KOLBAS}. 

One-dimensional models are also ideally suited to examine the mathematical 
foundations of Quantum Mechanics. In this paper, we shall take this 
foundational route. We shall construct the rigged Hilbert space (RHS) 
of the one-dimensional rectangular barrier potential, thereby showing that
the mathematical setting of quantum mechanical systems with continuous 
spectrum is the RHS rather than just the Hilbert space.

This paper follows up on Refs.~\cite{DIS,JPA,FP02}, where the RHSs of 3D 
spherical shell potentials were constructed~\cite{FNOTE1}, and on 
Ref.~\cite{IJTP}, where the RHS of the 3D free Hamiltonian was constructed. The
present paper complements Refs.~\cite{DIS,JPA,FP02,IJTP} in the following ways:
\begin{itemize}
\item[$\bullet$] We treat a truly 1D model on the full real line, rather than
the radial part of a 3D model.

\item[$\bullet$] We construct the RHS of the algebra generated by 
the position, momentum and energy observables, rather than just the RHS of the 
Hamiltonian. 

\item[$\bullet$] We construct not only the Dirac kets but also 
the Dirac bras, thereby showing even more clearly that the RHS fully 
implements Dirac's bra-ket formalism.
\end{itemize}

The model we consider in this paper is supposed to represent a spinless
particle moving in one dimension and impinging on a barrier. The relevant 
observables to this system are the position $Q$, the momentum $P$ and the 
Hamiltonian $H$. These observables are represented by the following 
differential operators:
\begin{eqnarray}
      Qf(x)=xf(x) \, , \label{fdopp}  \\
      Pf(x)=-\rmi \hbar \frac{\rmd}{\rmd x}f(x) \, ,  \\
      Hf(x)=-\frac{\hbar ^2}{2m}\frac{\rmd}{\rmd x^2}f(x)+V(x)f(x) \, , 
         \label{fdoph}
\end{eqnarray}
where
\begin{equation}
           V(x)=\left\{ \begin{array}{ll}
                                0   &-\infty <x<a  \\
                                V_0 &a<x<b  \\
                                0   &b<x<\infty 
                  \end{array} 
                 \right. 
	\label{sbpotential}
\end{equation}
is the 1D rectangular barrier potential. Formally, these observables satisfy 
the following commutation relations:
\begin{eqnarray}
      \left[ Q,P \right] =\rmi \hbar I \, , \label{cr1} \\
      \left[ H,Q \right] =- \frac{\rmi \hbar}{m} P \, ,   \\
      \left[ H,P \right] = \rmi \hbar \frac{\partial V(x)}{\partial x} \, .  
                                                \label{cr3}
\end{eqnarray}
The finite linear combinations of powers of $P$, $Q$ and $H$ constitute the
algebra of the 1D rectangular barrier potential. This algebra will be denoted 
by ${\cal A}$.

The differential operators (\ref{fdopp})-(\ref{fdoph}) induce three linear
operators on the Hilbert space $L^2(\mathbb R ,\rmd x)$. We shall denote their
Hilbert space domains by ${\cal D}(Q)$, ${\cal D}(P)$ and ${\cal D}(H)$. A 
major shortcoming of the Hilbert space is that ${\cal D}(P)$, 
${\cal D}(Q)$ and ${\cal D}(H)$ do not remain stable under the action of 
the operators of ${\cal A}$, which prevents algebraic operations (e.g., sums,
multiplications and commutation relations) of observables from being well 
defined on the whole Hilbert space. Further, the operators $P$, $Q$ and $H$
are unbounded, and hence discontinuous, with respect to the topology of the 
Hilbert space. Mathematically speaking, these are the reasons why we introduce 
a subdomain $\mathbf \Phi$ of the Hilbert space such that: 
\begin{itemize}
\item[({\it i})] The subdomain $\mathbf \Phi$ remains stable under the action 
of ${\cal A}$. This stability makes, in particular, algebraic operations such 
as the commutation relations~(\ref{cr1})-(\ref{cr3}) well defined on 
$\mathbf \Phi$.

\item[({\it ii})] The operators of ${\cal A}$ are continuous with respect to
a properly chosen topology of $\mathbf \Phi$.
\end{itemize}
As we shall see, the space $\mathbf \Phi$ is given by the maximal invariant 
subspace of ${\cal A}$.

The spectrum of $P$, $Q$ and $H$ is respectively $(-\infty ,\infty)$, 
$(-\infty ,\infty)$ and $[0,\infty)$. If ${\rm Sp}(A)$ denotes
the spectrum of the operator $A$, where $A$ can denote $P$, $Q$ or $H$, then
with each element $a$ of ${\rm Sp}(A)$ we associate a Dirac ket $|a\rangle$ and
a Dirac bra $\langle a|$ such that
\begin{itemize}
\item[({\it i})] The ket $|a\rangle$ is a right eigenvector of $A$ with 
eigenvalue $a$,
\begin{equation}
      A|a\rangle =a |a\rangle \, ,
       \label{keigen}
\end{equation}
and the bra $\langle a|$ is a left eigenvector of $A$ with eigenvalue $a$,
\begin{equation}
      \langle a|A =a \langle a|  \, .
         \label{beigen}
\end{equation}
\item[({\it ii})] The kets and bras are $\delta$-normalized,
\begin{equation}
      \langle a|a^{\prime}\rangle  = \delta (a-a^{\prime})  \, .
        \label{deltanormabk}
\end{equation}
\item[({\it iii})] The kets and bras form a complete basis system that can
be used to expand any wave function $\varphi$,
\begin{equation}
      \varphi = \sum_{\alpha} \int_{{\rm Sp}(A)} \rmd a \, 
          |a\rangle _{\alpha} \,  _{\alpha}\langle a|\varphi \rangle   \, ,
       \qquad A=P,Q,H \, ,
       \label{diracbve}
\end{equation}
where the label $\alpha$ accounts for any possible degeneracy of the spectrum.
\end{itemize} 
However, because the spectra of $P$, $Q$ and $H$ are continuous, the bras
and kets are not in the Hilbert space. Indeed, the Dirac kets $|a\rangle$ 
belong to the {\it antidual} space of $\mathbf \Phi$, which we shall denote by
$\mathbf \Phi ^{\times}$, whereas the Dirac bras $\langle a|$ belong to
the {\it dual} space of $\mathbf \Phi$, which we shall denote by
$\mathbf \Phi ^{\prime}$. Further, the Dirac basis vector expansions
(\ref{diracbve}) do not hold for all the elements of 
$\cal H$---Eq.~(\ref{diracbve}) holds only when $\varphi$ belongs to 
$\mathbf \Phi$.

We are thus led to two Gel'fand triplets
\begin{equation}
     \mathbf \Phi \subset {\cal H} \subset \mathbf \Phi ^{\times} 
\end{equation}
and
\begin{equation}
     \mathbf \Phi \subset {\cal H} \subset \mathbf \Phi ^{\prime} \, .
\end{equation}
The space $\mathbf \Phi$ contains those square integrable functions that can
be considered as physical, since algebraic operations (e.g., the commutation 
relations of observables) are well defined on $\mathbf \Phi$ but not the whole 
Hilbert space. The space $\mathbf \Phi ^{\times}$ contains the Dirac kets, 
that is, $\mathbf \Phi ^{\times}$ contains
the generalized right eigenvectors of the observables of the algebra. The 
space $\mathbf \Phi ^{\prime}$ contains the Dirac bras, that is, 
$\mathbf \Phi ^{\prime}$ contains
the generalized left eigenvectors of the observables of the 
algebra. Furthermore, the expansions (\ref{diracbve}) hold 
only when $\varphi$ is in $\mathbf \Phi$. Aside from providing us with 
mathematical concepts such as unitarity, self-adjointness and so on, the 
Hilbert space ${\cal H}$ singles out the scalar product that is used to 
calculate probability amplitudes. 

We recall that the RHS uses distribution theory to give meaning to the 
eigenvalue equations (\ref{keigen}) and (\ref{beigen}). Within the RHS, 
Eq.~(\ref{keigen}) means that
\begin{equation}
      \langle \varphi |A|a\rangle \equiv 
         \langle A^{\dagger}\varphi |a\rangle =
         a \langle \varphi |a\rangle \, , \qquad
        \forall \varphi \in \mathbf \Phi \, , 
       \label{keigenrhs}
\end{equation}
and Eq.~(\ref{beigen}) means that
\begin{equation}
      \langle a|A|\varphi \rangle \equiv 
      \langle a|A^{\dagger}\varphi \rangle =
           a \langle a| \varphi \rangle \, , \qquad
        \forall  \varphi \in \mathbf \Phi  \, .
         \label{beigenrhs}
\end{equation}
Sometimes, whenever is necessary to make clear that the operator $A$ in 
Eqs.~(\ref{keigenrhs}) and (\ref{beigenrhs}) is acting outside the Hilbert
space, we shall write these equations as
\begin{equation}
      \langle \varphi |A^{\times}|a\rangle =a \langle \varphi |a\rangle \, , 
           \qquad   \forall \varphi \in \mathbf \Phi \, , 
       \label{keigenrhsc}
\end{equation}
\begin{equation}
      \langle a|A^{\prime}|\varphi \rangle =a \langle a|\varphi \rangle \, , 
         \qquad  \forall  \varphi \in \mathbf \Phi  \, .
         \label{beigenrhsp}
\end{equation}
Here, $A^{\times}$ denotes the \emph{antidual} extension of $A$ acting to the 
right on the elements of $\mathbf \Phi ^{\times}$, whereas $A^{\prime}$ 
denotes the \emph{dual} extension of $A$ acting to the left on the elements of 
$\mathbf \Phi ^{\prime}$. Nevertheless, we shall normally use $A$ to denote
both $A^{\times}$ and $A^{\prime}$ unless there is a risk of confusion.

The ``scalar product'' $\langle a|a'\rangle =\delta (a-a')$ in 
Eq.~(\ref{deltanormabk}) should not be 
interpreted as an actual scalar product of two functionals $\langle a|$ and 
$|a'\rangle$. The delta function $\langle a|a'\rangle =\delta (a-a')$ appears 
as the kernel of the functionals $\langle a|$ and $|a'\rangle$ when we write
these functionals as integral operators. As well, the delta function 
$\delta (a-a')$ is the solution to the eigenvalue equation of the operator 
$A$ in the representation in which $A$ acts as multiplication by $a$. In 
general, given two operators $A$ and 
$B$, quantities such as $\langle b|a\rangle$ are distributions that are 
obtained by solving a differential eigenequation in the $b$-representation:
\begin{equation}
       \langle b|A|a\rangle =A \langle b|a \rangle =a \langle b|a\rangle \, .
\end{equation}
The $\langle b|a\rangle$ can also be seen as transition elements from the 
$a$- into the $b$-representation. Similar to the Dirac delta,  
$\langle b|a\rangle$ appears as the kernel of the functionals $\langle b|$ and 
$|a\rangle$ when we write these functionals as integral operators:
\begin{equation}
       \langle \varphi |a\rangle = \int \rmd b \, 
                    \langle \varphi |b\rangle \langle b|a\rangle \, , \quad
                     \varphi \in {\mathbf \Phi} \, ,
\end{equation}
\begin{equation}
       \langle \varphi |b\rangle = \int \rmd a \, 
                    \langle \varphi |a\rangle \langle a|b\rangle \, , \quad
                     \varphi \in {\mathbf \Phi} \, .
\end{equation}
In this paper, we shall 
encounter a few of these ``scalar products'' such as $\langle x| p\rangle$, 
$\langle x| x'\rangle$ and $\langle x| E^{\pm} \rangle _{\rm l,r}$, which
will be respectively obtained as the eigensolution to the eigenvalue equation 
for the operator $P$, $Q$ and $H$.

Fourier methods play a central role in any problem that is both linear and 
shift invariant, and this includes many wave-related 
phenomena~\cite{FORBES}. Hence, Fourier methods are at the foundations of the 
modeling of, for example, sound and light. As a result, the notions of 
frequency decomposition and uncertainty principle (that there is a minimum 
width to the spectrum that is inversely proportional to the localization of 
the signal) are widely used. Quantum Mechanics also uses Fourier methods. In
Quantum Mechanics, the Fourier transform relates the position and the energy 
representations, it amounts to a decomposition in plane waves, and it entails
an uncertainty principle. The difference between classical wave phenomena and
Quantum Mechanics is that, in the quantum case, what is ``waving'' is the
probability amplitude. Other than that, the analogy is very close.

In this paper, we shall push this analogy further and see that Dirac's 
formalism is, in fact, a generalization of Fourier methods: The classical 
monochromatic plane waves correspond to the Dirac bras and kets; the classical
frequency decomposition corresponds to the Dirac basis vector
expansions (\ref{diracbve}); the classical uncertainty principle of Fourier 
Optics corresponds to the quantum uncertainty principle associated with two 
non-commuting observables. Thus, in a way, Dirac's formalism is to Quantum 
Mechanics what Fourier methods are to classical wave-related phenomena.

The structure of the paper is the same as that in 
Refs.~\cite{DIS,JPA,FP02,IJTP}: In 
Sections~\ref{sec:domsadi}--\ref{sec:diagonalization}, we shall use the
Sturm-Liouville theory to obtain the self-adjoint extension of $H$ (in
Section~\ref{sec:domsadi}), the resolvent and the Green function of $H$
(in Section~\ref{sec:reopangreefu}), the spectrum of $H$ (in 
Section~\ref{sec:spectrum}), and the diagonalizations and eigenfunction 
expansions associated to $H$ (in Section~\ref{sec:diagonalization}). In
Section~\ref{sec:consrhs}, we shall construct the RHS, the bras, the kets
and the Dirac basis vector expansion of the 1D rectangular barrier. The
results of Section~\ref{sec:consrhs} will essentially follow from those in
Refs.~\cite{ROBERTSJMP,ROBERTSCMP,DIS}. At the end of 
Section~\ref{sec:consrhs}, we shall construct the energy, the momentum and the 
wave-number representations of the RHS of the 1D rectangular 
potential. Finally, the conclusions of the paper will be inserted in 
Section~\ref{sec:conclusions}. Throughout the paper, we shall recall the 
well-known results for $Q$, $P$ and the Fourier transform, in order to show 
that Dirac's formalism can be viewed as a Fourier-like formalism.

Before proceeding with the main body of the paper, we recall that the RHS
is becoming a standard research tool in many areas of theoretical physics,
especially in Quantum Mechanics. The RHS is widely used in the quantum theory 
of scattering and decay (see Refs.~\cite{DIS,BG,BOLLINI} and references 
therein), in 
the ongoing effort to construct quantum time operators~\cite{GALAPON}, and in 
the construction of generalized spectral decompositions of chaotic 
maps~\cite{AT93,SUCHANECKI}. We note, however, that the use some authors
make of the RHS differs from ours in many fundamental ways. For example, in 
Ref.~\cite{BOHM}, the authors claim that the Hilbert space 
is ``only of pedagogical or historical importance'' and therefore, according 
to A.~Bohm {\it et al.}, only the dual pair $\mathbf \Phi$, 
$\mathbf \Phi ^{\times}$ should matter (see Ref.~\cite{BOHM}, 
page~443). However, from the results of this paper, it should be clear that 
the RHS is an extension (rather than a replacement) of the Hilbert space, and 
that the RHS arises when we equip the Hilbert space with distribution 
theory. In particular, the Hilbert space is at the core of the RHS.

\section{Domains of self-adjointness and deficiency indexes}
\label{sec:domsadi}

In order to make the paper self-contained, we recall in this section the 
domains of the Hilbert space $L^2({\mathbb R}, \rmd x)$ on which $Q$, $P$ and 
$H$ are self-adjoint. In what follows, it will be convenient to 
distinguish between the formal differential operator (which can act inside 
and outside the Hilbert space) and the self-adjoint operator (which is 
defined by the formal differential operator and by the Hilbert space domain on 
which it acts). 

The formal differential operator corresponding to the energy observable will 
be denoted by $h$:
\begin{equation}
       h\equiv -\frac{\hbar ^2}{2m}\frac{\rmd ^2 \ }{\rmd x^2}+V(x) \, .
      \label{doh}
\end{equation}
In order to obtain the domain on which $h$ induces the self-adjoint operator
$H$, we need first some definitions (cf.~\cite{DUNFORDII}):

\vskip0.5cm

\theoremstyle{definition}
\newtheorem*{Def1}{Definition~1}
\begin{Def1} By $AC (\mathbb R)$ we denote the space of 
all functions $f$ which are not only continuous but also 
absolutely continuous over each compact subinterval of $\mathbb R$. Thus,
$f'$ exists almost everywhere and is integrable over any compact 
subinterval of $\mathbb R$. 

By $AC^2 (\mathbb R)$ we denote the space of 
all functions $f$ which have a continuous derivative in 
$\mathbb R$, and for which $f'$ is not only continuous but also 
absolutely continuous over each compact subinterval of $\mathbb R$. Thus,
$f ^{(2)}$ exists almost everywhere and is integrable over any compact 
subinterval of $\mathbb R$. 
\end{Def1}

\vskip0.5cm

The space $AC (\mathbb R)$ is the largest space of functions on which 
the differential operator $-\rmi \hbar \rmd /\rmd x$ can be defined. The space 
$AC ^2 (\mathbb R)$ is the largest space of functions on which the 
differential operator $h$ can be defined. 

\vskip0.5cm

\theoremstyle{definition}
\newtheorem*{Def2}{Definition~2}
\begin{Def2} We define the spaces
\begin{eqnarray}
 \hskip-1cm  &&
      {\cal H}^2 _h (\mathbb R)\equiv \{ f\in AC ^2 (\mathbb R) \, : \   
                                       f, hf \in L^2 (\mathbb R,\rmd x) \} \\
   \hskip-1cm &&
      {\cal H}^2  (\mathbb R)\equiv \{ f\in AC ^2 (\mathbb R) \, : \ f,
                                       f^{(2)} \in L^2 (\mathbb R,\rmd x) \} \\
   \hskip-1cm  &&
      {\cal H}^2 _{\rm min} (\mathbb R)\equiv \{ f\in {\cal H}^2(\mathbb R) 
         \, : \ f \  \mbox{vanishes outside some compact subset of} 
       \ \mathbb R \}  \, .
\end{eqnarray}
\end{Def2}

\vskip0.5cm

By using these spaces, we can define the necessary operators to obtain the 
self-adjoint extensions associated to $h$.

\vskip0.5cm

\theoremstyle{definition}
\newtheorem*{Def3}{Definition~3}
\begin{Def3} Let $h$ be the formal differential operator 
(\ref{doh}). The operators $H_{\rm min}$ and $H_{\rm max}$ are defined on 
$L^2 (\mathbb R,\rmd x)$ by the formulas
\begin{eqnarray}
      &&{\cal D}(H_{\rm min})={\cal H}^2 _{\rm min}(\mathbb R), 
                             \quad H_{\rm min}f:=hf, \quad
       f \in {\cal D}(H_{\rm min}) \, . \\
      &&{\cal D}(H_{\rm max})={\cal H}^2 _h (\mathbb R), \quad 
              H_{\rm max}f:=hf, \quad   f\in {\cal D}(H_{\rm max}) \, .    
\end{eqnarray} 
\end{Def3}

\vskip0.5cm

The operators $H_{\rm min}$ and $H_{\rm max}$ are sometimes called the 
{\it minimal} and the {\it maximal} operators associated to the differential 
operator $h$, respectively. The domain ${\cal D}(H_{\rm max})$ is the largest 
domain of $L^2 (\mathbb R,\rmd x )$ on which the action of $h$ can be defined 
and remains inside $L^2 (\mathbb R,\rmd x)$. Further, 
$H_{\rm min}^{\dagger}=H_{\rm max}$.

The self-adjoint extensions we are looking for are operators $H$ such that
\begin{equation}
     H_{\rm min} \subset H \subset H_{\rm max} \, .
\end{equation}
In order to obtain these self-adjoint extensions, we need to calculate the
deficiency indexes $n_{\pm}(H)$, which are the number of linearly independent
solutions of the equations
\begin{equation}
      H_{\rm min}^{\dagger}f=\pm \rmi \lambda f 
       \, , \quad f \in {\cal D}(H^{\dagger}_{\rm min})  \, ,
       \label{defininequs}
\end{equation}
where $\lambda >0$ has been introduced on dimensional grounds. It is 
straightforward to see that the only solution of 
Eq.~(\ref{defininequs}) that belongs to ${\cal D}(H^{\dagger}_{\rm min})$ is 
the zero solution, that is, $n_{\pm}(H)=0$. Thus, the only domain ${\cal D}(H)$
of $L^2 (\mathbb R,\rmd x )$ on which $h$ induces a self-adjoint operator
coincides with the maximal domain: 
\begin{equation}
     {\cal D}(H)=\left\{ f\in L^2(\mathbb R,\rmd x) \, : \ 
                f \in  AC^2(\mathbb R,\rmd x), \
                hf \in  L^2(\mathbb R,\rmd x) \right\} \, . 
         \label{domainH} 
\end{equation}

By similar arguments, it can be shown that the only domain on which the
multiplication operator induces a self-adjoint operator is given by
\begin{equation}
     {\cal D}(Q)=\left\{ f\in L^2(\mathbb R,\rmd x) \, : 
                         xf \in  L^2(\mathbb R,\rmd x) \right\} \, ,  
\end{equation}
and that the only domain on which the differential operator 
$-\rmi \hbar \rmd /\rmd x$ induces a self-adjoint operator is given by
\begin{equation}
     {\cal D}(P)=\left\{ f\in L^2(\mathbb R, \rmd x) \, : \ 
                f \in  AC(\mathbb R, \rmd x), \
                f' \in  L^2(\mathbb R,\rmd x) \right\} \, . 
\end{equation}

\section{The resolvent operator and the Green function}
\label{sec:reopangreefu}

In this section, we obtain the resolvent and the Green function of $H$, which
can be easily calculated by way of the following theorem 
(cf.~Theorem~XIII.3.16 of Ref.~\cite{DUNFORDII}):

\vskip0.5cm

\theoremstyle{plain}
\newtheorem*{Th1}{Theorem~1}
\begin{Th1} Let $H$ be the self-adjoint Hamiltonian operator 
derived from the real formal differential operator (\ref{doh}) and
the domain (\ref{domainH}). Let ${\rm Im}(E) \neq 0$. Then there is exactly 
one solution $\chi _{\rm r}(x;E)$ of $(h-E)\sigma =0$ square integrable at 
$-\infty$, and exactly one solution $\chi _{\rm l}(x;E)$ of $(h-E)\sigma =0$ 
square-integrable at $+\infty$. The resolvent $(E-H)^{-1}$ is an integral 
operator whose kernel $G(x,x';E)$ is given by
\begin{equation}
       G(x,x';E)=\left\{ \begin{array}{ll}
               \frac{2m}{\hbar ^2} \,
      \frac{\chi _{\rm r}(x;E) \, \chi _{\rm l}(x';E)}
             {W(\chi _{\rm r},\chi _{\rm l})}
               &x<x' \\ [1ex] 
      \frac{2m}{\hbar ^2} \,
      \frac{\chi _{\rm r}(x';E) \, \chi _{\rm l} (x;E)}
               {W(\chi _{\rm r},\chi _{\rm l} )}
                       &x>x'  \, ,
                  \end{array} 
                 \right. 
	\label{exofGFA}
\end{equation}
where $W(\chi _{\rm r},\chi _{\rm l} )$ is the Wronskian of $\chi _{\rm r}$
and $\chi _{\rm l}$:
\begin{equation}
       W(\chi _{\rm r},\chi _{\rm l} )=
       \chi _{\rm r}\chi _{\rm l}'-\chi _{\rm r}'\chi _{\rm l} \, .
\end{equation}
\end{Th1}

\vskip0.5cm

To obtain $G(x,x';E)$, we divide the complex $E$-plane in three 
regions (left half-plane, first quadrant, and fourth quadrant) and apply
Theorem~1 to each of these regions separately. In our calculations, we shall 
use the following branch of the square root function:
\begin{equation}
   \hskip-1.7cm   \sqrt{\cdot}:\{ E\in {\mathbb C} \, : \  
       -\pi <{\rm arg}(E)\leq \pi \} 
   \longmapsto \{E\in {\mathbb C} \, : \  -\pi/2 <{\rm arg}(E)\leq \pi/2 \} 
      \, .
   \label{branch} 
\end{equation}
This branch is chosen because it grants the following relation:
\begin{equation}
      \overline{\sqrt{\overline{z}\, }}=z \, , \quad z\in \mathbb C \, .
     \label{cczez}
\end{equation}

It is important to keep in mind that $\chi _{\rm r}(x;E)$ and 
$\chi _{\rm l}(x;E)$ of Theorem~1 are well defined for real as well as for 
complex energies. More precisely, the functions $\chi _{\rm r}(x;E)$ and 
$\chi _{\rm l}(x;E)$, which are derived for complex $E$ of
nonzero imaginary part, have a well-defined limiting value when $E$ approaches
the real line. The values of $\chi _{\rm r}(x;E)$ and 
$\chi _{\rm l}(x;E)$ for complex $E$ will be used in this section
to calculate $G(x,x';E)$. The values of $\chi _{\rm r}(x;E)$ and 
$\chi _{\rm l}(x;E)$ for real $E$ will be used in Sec.~\ref{sec:consrhs}
to construct the bras and kets associated to the energies in the spectrum 
of the Hamiltonian.

\subsection{Left half-plane: ${\rm Re}(E)<0$, ${\rm Im}(E)\neq 0$}

According Theorem~1, we need to obtain the eigensolutions 
$\widetilde{\chi}_{\rm r}$ and $\widetilde{\chi}_{\rm l}$ of the Schr\"odinger 
equation
\begin{equation}
      \left( -\frac{\hbar ^2}{2m}\frac{\rmd ^2 \ }{\rmd x^2}+V(x)\right) 
       \sigma (x;E)= E \sigma (x;E) 
      \label{sde}
\end{equation}
that are square integrable at $-\infty$ and at $\infty$, respectively. Thus, 
the eigensolution $\widetilde{\chi}_{\rm r}(x;E)$ satisfies
\numparts
\begin{eqnarray}
       && \widetilde{\chi}_{\rm r} (x;E)\in AC^2(\mathbb R) \, , 
                        \label{bcchiac} \\
       && \widetilde{\chi}_{\rm r}(x;E)
           {\rm \ is \ square \ integrable \ at \ } -\infty \, ,
       \label{sbca03} 
\end{eqnarray}
      \label{eigsbo0co}
\endnumparts
whereas the eigensolution $\widetilde{\chi}_{\rm l}(x;E)$ satisfies
\numparts
\begin{eqnarray}
       &&\widetilde{\chi}_{\rm l}(x;E)\in AC^2(\mathbb R) \, , 
                   \label{bcainfty1} \\
       &&\widetilde{\chi}_{\rm l}(x;E) \ 
         {\rm is \ square \ integrable \ at \ } +\infty \, . \label{bcainfty2}
\end{eqnarray}
      \label{thetcoabejej}
\endnumparts

Solving Eq.~(\ref{sde}) subjected to (\ref{bcchiac})-(\ref{sbca03}) yields
\begin{equation}
      \widetilde{\chi}_{\rm r}(x;E)=
          \left( \frac{m}{2\pi \widetilde{k} \hbar ^2} \right)^{1/2} \times
          \left\{ \begin{array}{lc}
              \widetilde{T}(\widetilde{k})\rme ^{\widetilde{k}x}
                 \quad &-\infty<x<a  \\
            \widetilde{A}_{\rm r}(\widetilde{k})
                  \rme ^{-\widetilde{Q}x}+
             \widetilde{B}_{\rm r}(\widetilde{k})
                      \rme ^{\widetilde{Q}x}
                 \quad  &a<x<b \\
           \widetilde{R}_{\rm r}(\widetilde{k})\rme ^{-\widetilde{k}x}+
           \rme ^{\widetilde{k}x}
                 \quad  &b<x<\infty \, ,
               \end{array} 
                 \right. 
       \label{tildechifunction}
\end{equation}
where
\begin{equation}
      \widetilde{k}=\sqrt{-\frac{2m}{\hbar ^2}E} \, , \quad
      \widetilde{Q}=\sqrt{-\frac{2m}{\hbar ^2}(E-V_0)} \, ,
\end{equation}
where the coefficients $\widetilde{T}(\widetilde{k})$,
$\widetilde{A}_{\rm r}(\widetilde{k})$, 
$\widetilde{B}_{\rm r}(\widetilde{k})$ and 
$\widetilde{R}_{\rm r}(\widetilde{k})$
can be found in \ref{sec:appauxfunc}, and where 
$\left( \frac{m}{2\pi \widetilde{k} \hbar ^2} \right)^{1/2}$ is a 
normalization factor.

Solving Eq.~(\ref{sde}) subjected 
to (\ref{bcainfty1})-(\ref{bcainfty2}) yields
\begin{equation}
      \widetilde{\chi}_{\rm l}(x;E)=
          \left( \frac{m}{2\pi \widetilde{k} \hbar ^2} \right)^{1/2} \times
          \left\{ \begin{array}{lc}
             \rme ^{-\widetilde{k}x}+ 
                \widetilde{R}_{\rm l}(\widetilde{k})\rme ^{\widetilde{k}x} 
                 \quad &-\infty<x<a  \\
            \widetilde{A}_{\rm l}(\widetilde{k})
              \rme ^{-\widetilde{Q}x}+    
             \widetilde{B}_{\rm l}(\widetilde{k})
               \rme ^{\widetilde{Q}x}   \quad  &a<x<b \\
             \widetilde{T}(\widetilde{k})
             \rme ^{-\widetilde{k}x}  \quad  &b<x<\infty \, ,
               \end{array} 
                 \right. 
       \label{tildethetfunc}
\end{equation}
where the coefficients $\widetilde{R}_{\rm l}(\widetilde{k})$,
$\widetilde{A}_{\rm l}(\widetilde{k})$, 
$\widetilde{B}_{\rm l}(\widetilde{k})$ and 
$\widetilde{T}(\widetilde{k})$ can be found in 
\ref{sec:appauxfunc}. 

The Wronskian of $\widetilde{\chi}_{\rm r}$ and 
$\widetilde{\chi}_{\rm l}$ is given by
\begin{equation}
   W(\widetilde{\chi}_{\rm r},\widetilde{\chi}_{\rm l})=-
    \frac{m}{\pi \hbar ^2} \,
        \widetilde{T}(\widetilde{k}) \, .
   \label{wronskiantilde}
\end{equation}

From Eqs.~(\ref{exofGFA}), (\ref{tildechifunction}), (\ref{tildethetfunc})
and (\ref{wronskiantilde}), it follows that, when $E$ belongs to the
left half-plane, the expression of the Green function reads as
\begin{equation}
     \hskip-1.5cm  G(x,x';E)=\left\{ \begin{array}{ll}
               -2\pi \,
    \frac{\widetilde{\chi}_{\rm r}(x;E) \, \widetilde{\chi}_{\rm l}(x';E)}
          {\widetilde{T}(E)}
               &x<x' \\ 
     -2\pi \,
    \frac{\widetilde{\chi}_{\rm r}(x';E) \, \widetilde{\chi}_{\rm l}(x;E)}
          {\widetilde{T}(E)}
                       &x>x'  
                  \end{array} 
                 \right. \quad \mbox{Re}(E)<0 \, , \ \mbox{Im}(E)\neq 0 \, .
      \label{green-}
\end{equation}

\subsection{First quadrant: ${\rm Re}(E)>0$, ${\rm Im}(E)> 0$}
\label{sec:region++}

When $E$ belongs to the first quadrant, the eigensolution 
${\chi}_{\rm r}^+$ that satisfies Eq.~(\ref{sde}) subjected to the boundary 
conditions (\ref{bcchiac})-(\ref{sbca03}) is given by
\begin{equation}
           \chi _{\rm r}^+(x;E)= 
          \left( \frac{m}{2\pi k \hbar ^2} \right)^{1/2} \times   
                \left\{ \begin{array}{lc}
             T (k)\rme ^{-\rmi kx}  &-\infty <x<a  \\
             A_{\rm r}(k)\rme ^{\rmi Qx}+B_{\rm r}(k)\rme ^{-\rmi Qx}&a<x<b \\
        R_{\rm r}(k)\rme ^{\rmi kx} + \rme ^{-\rmi kx}   &b<x<\infty \, ,
                  \end{array} 
                 \right. 
     \label{chir+}
\end{equation}
where
\begin{equation}
      k=\sqrt{\frac{2m}{\hbar ^2}E} \, , \quad
      Q=\sqrt{\frac{2m}{\hbar ^2}(E-V_0)} \, ,
\end{equation}
where the coefficients $T(k)$, $A_{\rm r}(k)$, $B_{\rm r}(k)$ and 
$R_{\rm r}(k)$ can be found in \ref{sec:appauxfunc}, and where
$\left( \frac{m}{2\pi k \hbar ^2} \right)^{1/2}$ is a normalization 
factor. We recall that $\chi _{\rm r}^+(x;E)$ corresponds to a plane wave that 
impinges the barrier from the right (hence the subscript r) and gets reflected 
to the right with probability amplitude $R_{\rm r}(k)$ and transmitted to the 
left with probability amplitude $T (k)$.

The eigensolution of (\ref{sde}) subjected to 
(\ref{bcainfty1})-(\ref{bcainfty2}) is
\begin{equation}
           \chi _{\rm l}^+(x;E)= \left( \frac{m}{2\pi k \hbar ^2} \right)^{1/2}
               \times  \left\{ \begin{array}{lc}
             \rme ^{\rmi kx}+R_{\rm l}(k)\rme ^{-\rmi kx}  &-\infty <x<a  \\
             A_{\rm l}(k)\rme ^{\rmi Qx}+B_{\rm l}(k)\rme ^{-\rmi Qx}&a<x<b \\
               T (k)\rme ^{\rmi kx}         &b<x<\infty \, ,
                  \end{array} 
                 \right.
          \label{chil+} 
\end{equation}
where the coefficients ${R}_{\rm l}(k)$, ${A}_{\rm l}(k)$, ${B}_{\rm l}(k)$ and
${T}(k)$ can be found in \ref{sec:appauxfunc}. We recall that 
$\chi _{\rm l}^+(x;E)$
corresponds to a plane wave that impinges the barrier from the left (hence
the subscript l) and gets reflected to the left with probability amplitude
$R_{\rm l}(k)$ and transmitted to the right with probability amplitude 
$T(k)$.

The Wronskian of ${\chi}_{\rm r}^+$ and 
${\chi}_{\rm l}^+$ is given by
\begin{equation}
   W({\chi}_{\rm r}^+,{\chi}_{\rm l}^+)=\frac{\rmi m}{\pi \hbar ^2}\, T(k) \, .
\end{equation}

Thus, when $E$ belongs to the first quadrant, the expression of the Green 
function is given by
\begin{equation}
     \hskip-1.5cm  G(x,x';E)=\left\{ \begin{array}{ll}
               \frac{2\pi}{\rmi} \,
    \frac{{\chi}_{\rm r}^+(x;E) \, {\chi}_{\rm l}^+(x';E)}
          {T(k)}
               &x<x' \\ 
     \frac{2\pi}{\rmi} \,
    \frac{{\chi}_{\rm r}^+(x';E) \, {\chi}_{\rm l}^+(x;E)}
          {T(k)}
                       &x>x'  
                  \end{array} 
                 \right. \quad \mbox{Re}(E)>0 \, , \ \mbox{Im}(E)>0 \, .
      \label{green++}
\end{equation}

\subsection{Fourth quadrant: ${\rm Re}(E)>0$, ${\rm Im}(E)< 0$}

When $E$ belongs to the fourth quadrant, the eigensolution ${\chi}_{\rm r}^-$ 
that satisfies Eq.~(\ref{sde}) subjected to the boundary conditions 
(\ref{bcchiac})-(\ref{sbca03}) is given by
\begin{equation}
           \chi _{\rm r}^-(x;E)= \left( \frac{m}{2\pi k \hbar ^2} \right)^{1/2}
              \times \left\{ \begin{array}{lc}
             T^*(k)\rme ^{\rmi kx}  &-\infty <x<a  \\
         A_{\rm r}^*(k)\rme ^{-\rmi Qx}+B_{\rm r}^*(k)\rme ^{\rmi Qx}&a<x<b \\
        R_{\rm r}^*(k)\rme ^{-\rmi kx} + \rme ^{\rmi kx}   &b<x<\infty \, ,
                  \end{array} 
                 \right. 
             \label{chir-}
\end{equation}
where the coefficients $T^*(k)$, $A_{\rm r}^*(k)$, $B_{\rm r}^*(k)$ and 
$R_{\rm r}^*(k)$ can be found in \ref{sec:appauxfunc}. This eigensolution
corresponds to two plane waves---one impinging the barrier from the left with 
probability amplitude $T^*(k)$ and another impinging the barrier from the
right with probability amplitude $R_{\rm r}^*(k)$---that combine in such a way 
as to produce an outgoing plane wave to the right.

The eigensolution of (\ref{sde}) subjected to (\ref{bcainfty1})-(\ref{bcainfty2})
reads as
\begin{equation}
           \chi _{\rm l}^-(x;E)= \left( \frac{m}{2\pi k \hbar ^2} \right)^{1/2}
              \times \left\{ \begin{array}{lc}
         \rme ^{-\rmi kx}+R_{\rm l}^*(k)\rme ^{\rmi kx}  &-\infty <x<a  \\
         A_{\rm l}^*(k)\rme ^{-\rmi Qx}+B_{\rm l}^*(k)\rme ^{\rmi Qx}&a<x<b \\
               T^*(k)\rme ^{-\rmi kx}         &b<x<\infty \, ,
                  \end{array} 
                 \right.
          \label{chil-} 
\end{equation}
where the functions ${R}_{\rm l}^*(k)$, ${A}_{\rm l}^*(k)$, ${B}_{\rm l}^*(k)$ 
and ${T}^*(k)$ can be found in \ref{sec:appauxfunc}. This eigensolution
corresponds to two plane waves---one impinging the barrier from left
with probability amplitude $R_{\rm l}^*(k)$ and another impinging the 
barrier from the right with probability amplitude $T^*(k)$---that combine in 
such a way as to produce an outgoing wave to the left. Clearly, both
$\chi _{\rm r}^-(x;E)$ and $\chi _{\rm l}^-(x;E)$ correspond to the {\it final}
condition of an outgoing plane wave propagating respectively to the right and 
to the left, as opposed to $\chi _{\rm r}^+(x;E)$ and $\chi _{\rm l}^+(x;E)$, 
which correspond to the {\it initial} condition of a plane wave 
that impinges the barrier respectively from the left and from the right. 

The Wronskian of ${\chi}_{\rm r}^-$ and ${\chi}_{\rm l}^-$ is given by
\begin{equation}
   W({\chi}_{\rm r}^-,{\chi}_{\rm l}^-)=- \frac{\rmi m}{\pi \hbar ^2} \, T^*(k)
         \, .
\end{equation}

Thus, when $E$ belongs to the fourth quadrant, the expression of the Green 
function is given by
\begin{equation}
     \hskip-1.5cm  G(x,x';E)=\left\{ \begin{array}{lc}
               -\frac{2\pi}{\rmi} \,
    \frac{{\chi}_{\rm r}^-(x;E) \, {\chi}_{\rm l}^-(x';E)}
          {T^*(k)}
               &x<x' \\ 
     -\frac{2\pi}{\rmi} \,
    \frac{{\chi}_{\rm r}^-(x';E) \, {\chi}_{\rm l}^-(x;E)}
          {T^*(k)}
                       &x>x'  
                  \end{array} 
                 \right. \quad \mbox{Re}(E)>0 \, , \ \mbox{Im}(E)< 0 \, .
      \label{green+-}
\end{equation}

To finish this section, we recall the resolvents of $Q$ and $P$, which are 
well known and can be calculated by similar arguments. The resolvent of $Q$ 
is an integral operator whose kernel is
\begin{equation}
       \langle x|\frac{1}{z-Q}|x'\rangle = \frac{1}{z-x} \delta (x-x') \, ,
         \quad z \in \mathbb{C}/\mathbb{R} \, , \
                                      x, x' \in \mathbb{R} \, .
\end{equation}
In the upper complex $p$-plane, the kernel of the resolvent of $P$ is given 
by~\cite{DUNFORDII}
\begin{equation}      
       \langle x|\frac{1}{p-P}|x'\rangle = \left\{ 
                  \begin{array}{cl} 
                     0  & x<x'  \\
                     \frac{1}{\rmi \hbar}\,
                   \rme^{\rmi p(x-x')/\hbar}
                         & x>x'\end{array} 
                    \right. \quad {\rm Im}(p)>0 \, , 
\end{equation}
whereas in the lower complex $p$-plane it is given by~\cite{DUNFORDII}
\begin{equation}      
       \langle x|\frac{1}{p-P}|x'\rangle = \left\{ 
                  \begin{array}{cl} 
                  - \frac{1}{\rmi \hbar}\,
                   \rme ^{\rmi p(x-x')/\hbar} & x<x' \\
                    0   &  x>x'  \end{array} 
                    \right. \quad {\rm Im}(p)<0 \, . 
\end{equation}

\section{Spectrum}
\label{sec:spectrum}

In this section, we obtain the spectrum of $H$, which we shall 
denote by $\mbox{Sp}(H)$. In order to obtain $\mbox{Sp}(H)$, we shall apply 
Theorem~3 below. Before stating Theorem~3, we need to state Theorem~2, which 
provides the unitary operators that diagonalize $H$ 
(cf.~Theorem XIII.5.13 of Ref.~\cite{DUNFORDII}):

\vskip0.5cm

\theoremstyle{plain}
\newtheorem*{Th2}{Theorem~2}
\begin{Th2} (Weyl-Kodaira) Let $h$ be the formally self-adjoint 
differential operator~(\ref{doh}). Let $H$ be the
corresponding 
self-adjoint Hamiltonian. Let $\Lambda$ be an open interval of the real axis, 
and suppose that there is given a set $\{ \sigma _1(x;E),\, \sigma _2(x;E)\}$ 
of functions, defined and continuous on $\mathbb R \times \Lambda$, such that 
for each fixed $E$ in $\Lambda$, $\{ \sigma _1(x;E),\, \sigma _2(x;E)\}$ forms
a basis for the space of solutions of $h\sigma =E\sigma$. Then there exists a 
positive $2\times 2$ matrix measure $\{ \varrho _{ij} \}$ defined on
$\Lambda$, such that the limit 
\begin{equation}
      (Uf)_i(E)=\lim_{c\to 0}\lim_{d\to \infty} 
        \left[ \int_c^d f(x) \overline{\sigma _i(x;E)}\rmd x \right]
\end{equation}
exists in the topology of $L^2(\Lambda ,\{ \varrho _{ij}\})$ for each $f$ in 
$L^2(\mathbb R,\rmd x)$ and defines an isometric isomorphism $U$ of 
${\sf E}(\Lambda )L^2(\mathbb R,\rmd x)$ onto 
$L^2(\Lambda ,\{ \varrho _{ij}\})$, where ${\sf E}(\Lambda )$ is the spectral 
projection associated with $\Lambda$.
\end{Th2}

\vskip0.5cm

The spectral measures $\{ \rho _{ij}\}$ are provided by the following theorem 
(cf.~Theorem XIII.5.18 of Ref.~\cite{DUNFORDII}):

\vskip0.5cm

\theoremstyle{plain}
\newtheorem*{Th3}{Theorem~3}
\begin{Th3} (Titchmarsh-Kodaira) Let $\Lambda$ be an open interval 
of the real axis and $O$ be an open set in the complex plane containing 
$\Lambda$. Let ${\rm re}(H)$ be the resolvent set of $H$. Let 
$\{ \sigma _1(x;E),\, \sigma _2(x;E)\}$ be a set of functions 
which form a basis for the solutions of the equation $h\sigma =E\sigma$, 
$E\in O$, and which are continuous on $\mathbb R \times O$ and analytically 
dependent on $E$ for $E$ in $O$. Suppose that the kernel $G(x,x';E)$ for the 
resolvent $(E-H)^{-1}$ has a representation
\begin{equation}
      G(x,x';E)=\left\{ \begin{array}{lll}
                   \sum_{i,j=1}^2 \theta _{ij}^-(E)\sigma _i(x;E)
                   \overline{\sigma _j(x';\overline{E})}\, , &
                     \qquad & x<x' \, ,  \\ [1ex]
                  \sum_{i,j=1}^2 \theta _{ij}^+(E)\sigma _i(x;E)
                   \overline{\sigma _j(x';\overline{E})} \, ,&\qquad & x>x' 
                           \, ,
                  \end{array}
                 \right.
        \label{greenfunitthes}
\end{equation}
for all $E$ in ${\rm re}(H)\cap O$, and that $\{ \varrho _{ij} \}$ is a 
positive matrix measure on $\Lambda$ associated with $H$ as in Theorem 2. 
Then the functions $\theta _{ij}^{\pm}$ are analytic in ${\rm re}(H)\cap O$,
and given any bounded open interval $(E_1,E_2)\subset \Lambda$, we have for 
$1\leq i,j\leq 2$,
\begin{equation}
       \begin{array}{lll}
     \hskip-1cm 
       \varrho _{ij}((E_1,E_2))&=& \lim_{\delta \to 0}\lim_{\varepsilon \to 0+}
         \frac{1}{2\pi \rmi}\int_{E_1+\delta}^{E_2-\delta}
          [ \theta _{ij}^-(E-\rmi \varepsilon )
            -\theta _{ij}^-(E+\rmi \varepsilon )
          ]\rmd E \\ [1ex]
      \hskip-1cm \quad &=& \lim_{\delta \to 0}\lim_{\varepsilon \to 0+}
         \frac{1}{2\pi \rmi}\int_{E_1+\delta}^{E_2-\delta}
          [ \theta _{ij}^+(E-\rmi \varepsilon )-
            \theta _{ij}^+(E+\rmi \varepsilon ) ] \rmd E \, .
         \end{array}
      \label{specmesa} 
\end{equation}
\end{Th3}

\vskip0.5cm

From Eq.~(\ref{specmesa}), it is clear that in order to obtain 
$\mbox{Sp}(H)$ and $\varrho _{ij}$, we need to see on what real $E$'s the 
functions $\theta _{ij}^{\pm}(E)$ fail to be analytic. We shall do so by
taking $\Lambda$ in Theorem~3 to be $(-\infty ,0)$ and $(0,\infty )$.

\subsection{Negative Energy Real Line: $\Lambda =(-\infty ,0)$}
\label{sec:NeERlin}

On the negative real line, we choose the basis $\{ \sigma _1,  \sigma _2\}$
of Theorem~3 as
\numparts
\begin{eqnarray}
        &&\sigma _1(x;E)=\widetilde{\chi}_{\rm l}(x;E) \, ,  \label{tildsi1} \\
        &&\sigma _2(x;E)=\widetilde{\chi}_{\rm r}(x;E) \, . \label{tildsi2}
\end{eqnarray}
\endnumparts
From Eqs.~(\ref{cczez}) and (\ref{tildethetfunc}) it follows that
\begin{equation}
      \overline{\widetilde{\chi}_l(x';\overline{E})}=
      \widetilde{\chi}_l(x';E) \, .
       \label{croscctil}
\end{equation}
Now, by taking advantage of Eqs.~(\ref{tildsi1}), (\ref{tildsi2}) and 
(\ref{croscctil}), we write Eq.~(\ref{green-}) as
\begin{equation}
      \hskip-1.7cm G(x,x';E)= 
      -2\pi \, 
      \frac{\sigma _2(x;E) \overline{\sigma _1(x';\overline{E})}}
        {\widetilde{T}(E)}
              \, , \quad  
       x<x' \, , \  \mbox{Re}(E)<0 \, , \mbox{Im}(E) \neq 0 \, .
        \label{finag-retoco} 
\end{equation}
By comparing Eqs.~(\ref{greenfunitthes}) and (\ref{finag-retoco}) we see that
\begin{equation}
      \theta _{ij}^-(E)= \left(  \begin{array}{cc}
        0 & 0 \\
       -\frac{2\pi}{\widetilde{T}(E)}  & 0
                                \end{array}
        \right) \, , \quad \mbox{Re}(E)<0 \, , \  \mbox{Im}(E) \neq 0 \, .
\end{equation}
The functions $\theta _{ij}^-(E)$ are analytic in a neighborhood of 
$\Lambda =(-\infty ,0)$ except at the energies for which
${\widetilde{T}(E)}$ vanishes. Because for the 
potential~(\ref{sbpotential}) the function
\begin{eqnarray}
       \widetilde{T}(\widetilde{k})=\widetilde{T}(-\rmi k)=T(k)  
\end{eqnarray}
does not vanish on the positive $k$-imaginary axis (i.e., on the negative real 
axis), the segment $(-\infty ,0)$ belongs to the resolvent of $H$. 

We note in passing that if there had been bound states, the basis
(\ref{tildsi1}) and (\ref{tildsi2}) would not had been analytic, and we would 
have had to multiply $\widetilde{\chi}_{\rm l,r}$ by the denominator of
$\widetilde{T}$ in order to have an analytic basis.

\subsection{Positive Energy Real Line: $\Lambda =(0, \infty )$}
\label{sec:PeERlin}

On the positive real line, we choose the basis $\{ \sigma _1 , \sigma _2\}$
of Theorem~3 as
\numparts
\begin{eqnarray}
      &&\sigma _1(x;E)=\chi _{\rm l}^+(x;E)\, ,
      \label{sigma1=sci} \\ 
      &&\sigma _2(x;E)= \chi _{\rm r}^+(x;E)\, .
       \label{sigam2cos}  
\end{eqnarray}
\endnumparts
It is easy to see that Eqs.~(\ref{cczez}), (\ref{chir+}), (\ref{chil+}), 
(\ref{chir-}) and (\ref{chil-}) imply that
\numparts
\begin{eqnarray}
      &&\overline{\chi _{\rm l}^+(x;\overline{E})}=\chi _{\rm l}^-(x;E)\, ,
      \label{comcoml+l-} \\ 
      &&\overline{\chi _{\rm r}^+(x;\overline{E})}=\chi _{\rm r}^-(x;E)\, .
      \label{comcomr+r-} 
\end{eqnarray}
\endnumparts
Then, Eqs.~(\ref{chir+}), (\ref{chil-}) and 
(\ref{sigma1=sci})-(\ref{comcomr+r-}) lead to
\numparts
\begin{eqnarray}
    &&\chi _{\rm r}^+(x';E)=T(E)\overline{\sigma _1(x';\overline{E})}-
    \frac{T(E)R_{\rm l}^*(E)}{T^*(E)}\overline{\sigma _2(x';\overline{E})}\, ,
      \label{chir+s1s2} \\
      &&\chi _{\rm l}^-(x;E)=-\frac{T^*(E)R_{\rm r}(E)}{T(E)}\sigma _1(x;E)
    +T^*(E)\sigma _2(x;E) \, , 
      \label{chil-s1s2} 
\end{eqnarray}
\endnumparts
After substituting Eq.~(\ref{chir+s1s2}) into Eq.~(\ref{green++}) and after 
some calculations, we get to
\begin{eqnarray}
     \hskip-0.6cm G(x,x';E)=
       \frac{2\pi}{\rmi} \,
       \left[ \sigma _1 (x;E) \overline{\sigma _1(x';\overline{E})}-
    \frac{R_{\rm l}^*(E)}{T^*(E)} \sigma _1 (x;E) 
   \overline{\sigma _2(x';\overline{E})} \right]  \, , \nonumber \\
      \qquad \hskip5.2cm  \mbox{Re}(E)>0, \mbox{Im}(E)>0\, , \, x>x' \, . 
        \label{redaot++}
\end{eqnarray}
After substituting Eq.~(\ref{chil-s1s2}) into Eq.~(\ref{green+-}) and after
some calculations, we get to
\begin{eqnarray}
      && \hskip-0.4cm G(x,x';E)=
       \frac{2\pi}{\rmi} \,
       \left[ \frac{R_{\rm r}(E)}{T(E)}\sigma _1(x;E)
         \overline{\sigma _2 (x';\overline{E})}
     -\sigma _2(x;E) \overline{\sigma _2 (x';\overline{E})} \right]      
       \, , \nonumber \\
      &&\qquad \hskip5.3cm \mbox{Re}(E)>0, \mbox{Im}(E)<0\, , \, x>x' \, .
      \label{redaot+-}
\end{eqnarray} 
By comparing (\ref{greenfunitthes}) to (\ref{redaot++}) we obtain
\begin{equation}
      \theta _{ij}^+(E)= \left(  \begin{array}{cc}
        \frac{2\pi}{\rmi}
       & -\frac{2\pi}{\rmi} \frac{R_{\rm l}^*(E)}{T^*(E)}\\
    0  & 0
                                \end{array}
        \right) 
        , \quad  \mbox{Re}(E)>0 \, , \  \mbox{Im}(E)>0 \, .
       \label{theta++}
\end{equation}
By comparing (\ref{greenfunitthes}) to (\ref{redaot+-}) we obtain
\begin{equation}
      \theta _{ij}^+(E)= \left(  \begin{array}{cc}
      0  & \frac{2\pi}{\rmi} \frac{R_{\rm r}(E)}{T(E)} \\
      0  & -\frac{2\pi}{\rmi}
                                \end{array}
        \right) 
        , \quad  \mbox{Re}(E)>0 \, , \  \mbox{Im}(E)<0 \, .
      \label{theta+-}
\end{equation}
Substitution of Eqs.~(\ref{theta++}) and (\ref{theta+-}) into 
Eq.~(\ref{specmesa})
yield the spectral measures $\varrho _{ij}$ of Theorem~3. The measure 
$\varrho _{21}$ is clearly zero. So is the measure $\varrho _{12}$, since
\begin{eqnarray}
       \varrho _{12}((E_1,E_2))&=&
       \lim _{\delta \to 0} \lim _{\varepsilon \to 0+}
      \frac{1}{2\pi \rmi} \int_{E_1+\delta}^{E_2-\delta}
      \left[ \theta _{12}^+ (E-\rmi \varepsilon ) 
             -\theta _{12}^+ (E+\rmi \varepsilon )
      \right] \rmd E \nonumber \\
      &=&\int_{E_1}^{E_2}
       - \left( \frac{R_{\rm r}(E)}{T(E)}+\frac{R_{\rm l}^*(E)}{T^*(E)}\right) 
           \rmd E \nonumber \\
      &=&0  \, ,
\end{eqnarray}
where in the last step we have used the relation
\begin{equation}
      R_{\rm r}(E)T^*(E)+T(E)R_{\rm l}^*(E)=0 \, .
\end{equation}
The measures $\varrho _{11}$ and $\varrho _{22}$ are just the Lebesgue measure,
since
\begin{eqnarray}
       \varrho _{11}((E_1,E_2))&=&
        \lim _{\delta \to 0} \lim _{\varepsilon \to 0+}
      \frac{1}{2\pi \rmi} \int_{E_1+\delta}^{E_2-\delta}
      \left[ \theta _{11}^+ (E-\rmi \varepsilon ) 
            -\theta _{11}^+ (E+\rmi \varepsilon )
      \right] \rmd E \nonumber \\
      &=&\int_{E_1}^{E_2} \rmd E =E_2-E_1 \, ,
\end{eqnarray}
and 
\begin{eqnarray}
       \varrho _{22}((E_1,E_2))&=&
        \lim _{\delta \to 0} \lim _{\varepsilon \to 0+}
      \frac{1}{2\pi \rmi} \int_{E_1+\delta}^{E_2-\delta}
      \left[ \theta _{22}^+ (E-\rmi \varepsilon ) 
         -\theta _{22}^+ (E+\rmi \varepsilon )
      \right] \rmd E \nonumber \\
      &=&\int_{E_1}^{E_2}  \rmd E =E_2-E_1 \, .
\end{eqnarray}
Clearly, the functions $\theta _{11}^+(E)$ and $\theta _{22}^+(E)$ both have a
branch cut along $(0,\infty)$, and therefore $(0,\infty )$ is included in 
${\rm Sp}(H)$. Since ${\rm Sp}(H)$ is a closed set, it must hold that
\begin{equation}
      \mbox{Sp}(H)= [0,\infty ) \, .
\end{equation}

We note that, instead of the ``initial'' basis 
(\ref{sigma1=sci})-(\ref{sigam2cos}), we could use the ``final'' basis:
\numparts
\begin{eqnarray}
      &&\sigma _1(x;E)=\chi _{\rm l}^-(x;E)\, , \\ 
      &&\sigma _2(x;E)= \chi _{\rm r}^-(x;E)\, .
\end{eqnarray}
\endnumparts
This basis produces the same spectrum (as it should be) and the
same spectral measures (cf.~\ref{sec:compfor-}).

To finish this section, we recall that the spectra of the position and
momentum observables coincide with the full real line:
\begin{equation}
      {\rm Sp}(Q)= {\rm Sp}(P) = (-\infty , \infty ) \, .
\end{equation}
The spectra of $Q$ and $P$ are simple, whereas the spectrum of $H$ is 
doubly degenerate. Indeed, to each energy $E\in [0,\infty )$ there correspond
two linearly independent eigenfunctions, $\chi _{\rm l}^+$ and 
$\chi _{\rm r}^+$ (or, equivalently, $\chi _{\rm l}^-$ and $\chi _{\rm r}^-$).

\section{Diagonalization and eigenfunction expansion}
\label{sec:diagonalization}

Theorem~2 of Sec.~\ref{sec:spectrum} provides the means to construct
two unitary operators $U_{\pm}$ that diagonalize $H$. The operator $U_+$
is associated with the basis $\{ \chi _{\rm l}^+, \chi _{\rm r}^+ \}$, 
whereas $U_-$ is associated with 
$\{ \chi _{\rm l}^-, \chi _{\rm r}^- \}$. These unitary operators transform
from the position into the energy representation, and they induce two
eigenfunction expansions and two direct integral decompositions of the Hilbert
space.

For the sake of brevity, we shall present each calculation associated with
$U_+$ together with the corresponding calculation associated with $U_-$.

By Theorem~2, the mappings $U_{\pm}$ are given by
\begin{equation}
\begin{array}{rcl}
   \hskip-0.5cm  
   U_{\pm}: L^2(\mathbb R, \rmd x)
      &\longmapsto & L^2([0,\infty),\rmd E)\oplus L^2( [0,\infty ),\rmd E)
                                        \nonumber \\
    \hskip-2cm   f(x) &\longmapsto & \widehat{f}^{\pm}(E)\equiv U_{\pm}f(E)
      \equiv
        \left[ (U_{\pm}f)_{\rm l}(E),(U_{\pm}f)_{\rm r}(E) \right] \, ,
\end{array}
    \label{operatoruplu}
\end{equation}
where
\begin{eqnarray}
    \widehat{f}_{\rm l}^{\pm}(E)\equiv (U_{\pm}f)_{\rm l}(E)=
         \int_{-\infty}^{\infty}\rmd x \, 
         f(x) \overline{\chi _{\rm l}^{\pm}(x;E)} \, , \label{hatEl+} \\
    \widehat{f}_{\rm r}^{\pm}(E)\equiv (U_{\pm}f)_{\rm r}(E)=
              \int_{-\infty}^{\infty}\rmd x \, 
         f(x) \overline{\chi _{\rm r}^{\pm}(x;E)} \, . \label{hatEr+}
\end{eqnarray}
Note that $\widehat{f}^{\pm}(E)\equiv U_{\pm}f(E)$ are two-component 
vectors, because the spectrum of $H$ is doubly degenerate. 

The inverses of $U_{\pm}$ can be obtained from the following theorem 
(cf.~Theorem XIII.5.14 of Ref.~\cite{DUNFORDII}):

\vskip0.5cm

\theoremstyle{plain}
\newtheorem*{Th4}{Theorem~4}
\begin{Th4} (Weyl-Kodaira) Let $H$, $\Lambda$, 
$\{ \varrho _{ij} \}$, etc., be as in Theorem~2. Let $E_0$ and $E_1$ be the 
end points of $\Lambda$. Then the inverse of the isometric isomorphism $U$ of 
${\sf E}(\Lambda )L^2(\mathbb R,\rmd x)$ onto 
$L^2(\Lambda ,\{ \varrho _{ij}\})$ is given by the formula
\begin{equation}
       (U^{-1}F)(x)=\lim_{\mu _0 \to E_0}\lim_{\mu _1 \to E_1}
       \int_{\mu _0}^{\mu _1} \left( \sum_{i,j=1}^{2}
             F_i(E)\sigma _j(x;E)\varrho _{ij}(\rmd E) \right)
\end{equation}
where $F=[F_1,F_2]\in L^2(\Lambda ,\{ \varrho _{ij}\})$, the limit existing
in the topology of $L^2(\mathbb R,\rmd x)$.
\end{Th4}

\vskip0.5cm

By Theorem~4, the inverses of $U_{\pm}$ are given by 
\begin{equation}
    \hskip-1cm f(x)=(U_{\pm}^{-1}\widehat{f})(x)= 
    \int_{0}^{\infty}\rmd E\, \widehat{f}_{\rm l}^{\pm}(E)
        \chi _{\rm l}^{\pm}(x;E) 
    +\int_{0}^{\infty}\rmd E\, \widehat{f}_{\rm r}^{\pm}(E)
        \chi _{\rm r}^{\pm}(x;E) \, ,
            \label{invdiagonaliza}
\end{equation}
where
\begin{equation}
     \widehat{f}_{\rm l}^{\pm}(E), \ \widehat{f}_{\rm r}^{\pm}(E)
     \in L^2([0,\infty ),\rmd E) \, .
\end{equation}
The operators $U_{\pm}^{-1}$ transform
from $L^2([0,\infty ),\rmd E)\oplus L^2([0,\infty ),\rmd E)$ onto 
$L^2(\mathbb R,\rmd x)$. Note that Eq.~(\ref{invdiagonaliza}) can also be
seen as the eigenfunction expansions of any element $f(x)$ of 
$L^2(\mathbb R, \rmd x )$ in terms of the basis 
$\{ \chi _{\rm l}^{\pm}, \chi _{\rm r}^{\pm} \}$. Similarly, we can write 
the two-component vectors $U_{\pm}f$, Eq.~(\ref{operatoruplu}), as
\begin{equation}
    \widehat{f}^{\pm}(E)\equiv \int_{-\infty}^{\infty}\rmd x \, 
         f(x) \overline{\chi _{\rm l}^{\pm}(x;E)} \dotplus
              \int_{-\infty}^{\infty}\rmd x \, 
         f(x) \overline{\chi _{\rm r}^{\pm}(x;E)} \, ,
     \label{eigenfucfinphi}
\end{equation}
which provide the eigenfunction expansions of any element 
$\widehat{f}^{\pm}(E)$
of $L^2([0,\infty ),\rmd E )\oplus L^2([0,\infty ),\rmd E)$ in terms of 
$\{ \chi _{\rm l}^{\pm}, \chi _{\rm r}^{\pm} \}$. (The symbol
$\dotplus$ in 
Eq.~(\ref{eigenfucfinphi}) intends to mean that, from a mathematical point of 
view, this equation should be seen as a two-component vector equality rather 
than as an actual sum.)

A straightforward calculation shows that
\begin{equation}
      \hskip-1.5cm \widehat{H}\widehat{f}^{\pm}(E)=
       U_{\pm}HU_{\pm}^{-1}\widehat{f}^{\pm}(E)=E\widehat{f}^{\pm}(E) \equiv
       [ E\widehat{f}_{\rm l}^{\pm}(E), E\widehat{f}_{\rm r}^{\pm}(E) ] \, , 
       \quad f\in {\cal D}(H) \, .
        \label{idagoplofh}
\end{equation}

The direct integral decompositions of the Hilbert space
induced by $U_{\pm}$ read as
\begin{equation}
\begin{array}{rcl}
      U_{\pm}:\mathcal{H}    &\longmapsto & 
      \widehat{\cal H}=
      \int_{0}^{\infty} \widehat{\cal H}_{\rm l}(E)\rmd E
      \oplus \int_{0}^{\infty} \widehat{\cal H}_{\rm r}(E)\rmd E \\
       f &\longmapsto & U_{\pm}f:= [(U_{\pm}f)_{\rm l},(U_{\pm}f)_{\rm r}] \, ,
\end{array}
   \label{dirintdec}
\end{equation}
where ${\cal H}$ is realized by $L^2(\mathbb R,\rmd x)$, and 
$\widehat{\cal H}$ is realized by
$L^2([0,\infty ),\rmd E)\oplus L^2([0,\infty ),\rmd E)$. The Hilbert spaces 
$\widehat{\mathcal H}_{\rm l}(E)$ and $\widehat{\mathcal H}_{\rm r}(E)$, which
are associated to each energy $E$ in the spectrum of $H$, are realized by 
$\mathbb C$. The scalar product on $\widehat{\cal H}$ can be written as
\begin{equation}
       \left( \widehat{f},\widehat{g} \right) _{\widehat{\cal H}}=
       \int_0^{\infty}\rmd E\,  
       \overline{\widehat{f}_{\rm l}^{\pm}(E)} \, \widehat{g}_{\rm l}^{\pm}(E)
         +\int_0^{\infty}\rmd E\,  
       \overline{\widehat{f}_{\rm r}^{\pm}(E)} \, \widehat{g}_{\rm r}^{\pm}(E)
        \, .
\end{equation}

It is worthwhile noticing the similarities between $U_{\pm}$ and the Fourier
transform $\cal F$, which is given by
\begin{equation}
\begin{array}{rcl}
      {\cal F}: L^2({\mathbb R},\rmd x)  &\longmapsto & 
      L^2({\mathbb R},\rmd p) \\
       f(x) &\longmapsto & {\cal F}f(p)=\frac{1}{\sqrt{2\pi \hbar}} 
       \int_{-\infty}^{\infty}\rmd x \, 
          f(x) \rme ^{-\rmi px/\hbar}    \, .
\end{array}
   \label{fourt}
\end{equation}
The operators $U_{\pm}$ transform between the position and the energy
representations, and $\cal F$ transforms between the position and the
momentum representations. The kernels of $U_{\pm}$, $\chi ^{\pm}_{\rm l,r}$, 
are eigenfunctions of the energy operator, and the kernel of $\cal F$,
$\frac{1}{\sqrt{2\pi \hbar}} \rme ^{-\rmi px/\hbar}$, is an eigenfunction of
the momentum operator. Like $\cal F$, $U_{\pm}$ are unitary operators. Thus,
$U_{\pm}$ are Fourier-like transforms.

\section{The rigged Hilbert space and Dirac's formalism}
\label{sec:consrhs}

In the previous sections, we have exhausted the Sturm-Liouville theory
(i.e., the Hilbert space mathematics) when applied to the rectangular
barrier. In this section, we equip the Sturm-Liouville theory with 
distribution theory, thereby constructing the equipped (i.e., rigged)
Hilbert space of the rectangular barrier.

\subsection{Construction of the rigged Hilbert space}
\label{sec:constr}

As explained in the Introduction, we need a subdomain of the Hilbert
space on which algebraic operations (sums, multiplications and commutation 
relations) involving $P$, $Q$ and $H$  are well defined and on which 
expectation values are finite. Essentially, that subdomain should remain 
stable under the action of any algebraic operation involving $P$, $Q$ and 
$H$. The largest of such subdomains is the maximal invariant subspace of the 
algebra $\cal A$~\cite{ROBERTSJMP,ROBERTSCMP}, which we shall denote by 
$\cal D$. Clearly, the elements 
of ${\cal D}$ must fulfill the following conditions:
\begin{itemize}
\item[$\bullet$] they are infinitely differentiable, so the differentiation
operation can be applied as many times as wished,
\item[$\bullet$] they vanish at $x=a$ and $x=b$, so differentiation is 
meaningful at the discontinuities of the potential~\cite{ROBERTSCMP},
\item[$\bullet$] the action of any power of the multiplication operator, of 
the differentiation operator and of $h$ is square integrable.
\end{itemize}
Hence,
\begin{eqnarray}
   \hskip-1cm {\cal D} =\{ \varphi \in L^2 (\mathbb R, \rmd x) \, : \
    \varphi \in C^{\infty}(\mathbb R), \ \varphi ^{(n)}(a)=\varphi ^{(n)}(b)=0 
    \, , \ n=0,1,\ldots \, ,  \nonumber \\
    \hskip3.6cm
     \frac{\rmd ^n}{\rmd x^n}x^mh^l\varphi (x) \in L^2 (\mathbb R, \rmd x) 
    \, , \   n,m,l=0,1, \ldots  \} \, .
     \label{ddomain}
\end{eqnarray}

The algebra of observables induces a natural topology on $\cal D$, whose
definition of convergence is as follows:
\begin{equation}
      \varphi _{\alpha}\, 
       \mapupdown{\tau_{\mathbf \Phi}}{\alpha \to \infty}
      \, \varphi \quad {\rm iff} \quad  
      \| \varphi _{\alpha }-\varphi \| _{n,m,l} 
      \, \mapupdown{}{\alpha \to \infty}\, 0 \, , \quad n,m,l=0,1, \ldots \, ,
\end{equation} 
where the norms $\| \, \cdot \, \|_{n,m,l}$ are defined as
\begin{equation}
      \| \varphi \| _{n,m,l} := 
 \sqrt{\int_{-\infty}^{\infty}\rmd x \, \left| P^nQ^mH^l\varphi (x)\right| ^2}
    \, , \quad n,m,l=0,1,\ldots \, .
      \label{nmnorms}
\end{equation}
When the space $\cal D$ is topologized by these norms, we obtain the locally
convex space of test functions $\mathbf \Phi$. On $\mathbf \Phi$, 
the expectation values
\begin{equation}
      (\varphi , A^n\varphi ) \, , \quad \varphi \in \mathbf \Phi \, ,
      \ A=P, Q, H 
\end{equation}
are finite, and the commutation relations (\ref{cr1})-(\ref{cr3}) are well 
defined. (Note that, when acting on $\varphi \in {\mathbf \Phi}$, the
commutation relation~(\ref{cr3}) becomes $[H,P]=0$, due to the vanishing
of the derivatives of $\varphi$ at the discontinuities of the potential.)
Moreover, the restrictions of $P$, $Q$ and $H$ to 
$\mathbf \Phi$ are essentially self-adjoint, $\tau _{\mathbf \Phi}$-continuous
operators. Equations~(\ref{ddomain}) and (\ref{nmnorms}) show
that $\mathbf \Phi$ is very similar to the Schwartz space, the
major differences being that the derivatives of the elements of $\mathbf \Phi$
vanish at $x=a,b$ and that $\mathbf \Phi$ is invariant not only under
$P$ and $Q$ but also under $H$. This is why we shall write
\begin{equation}
       \mathbf \Phi \equiv \Sw \, .
\end{equation}

Once we have constructed the space $\mathbf \Phi$, we can construct 
its topological antidual $\mathbf \Phi ^{\times}$ as the space of 
$\tau _{\mathbf \Phi}$-continuous {\it antilinear} functionals on 
$\mathbf \Phi$, and therewith the RHS corresponding to the algebra of the
1D rectangular barrier potential,
\begin{equation}
       \mathbf \Phi \subset {\cal H}\subset \mathbf \Phi ^{\times}
        \, ,
       \label{RHSCONT}
\end{equation}
which in the position representation is realized by
\begin{equation}
      \rhsSwt     \, .
       \label{RHSCONTpr}
\end{equation}

The space $\mathbf \Phi ^{\times}$ is meant to contain the eigenkets
$|p \rangle$, $|x\rangle$ and $|E^{\pm}\rangle _{\rm l,r}$ of $P$, $Q$ and
$H$. The definition of these eigenkets is borrowed from the theory of
distributions. The eigenket $|p\rangle$ is defined as an integral operator 
whose kernel is the eigenfunction of the differential operator 
$-\rmi \hbar \rmd /\rmd x$ with eigenvalue $p$:
\begin{equation}
\begin{array}{rcl}
       |p\rangle :\mathbf \Phi & \longmapsto & {\mathbb C} \\
       \varphi & \longmapsto & \langle \varphi |p\rangle  := 
       \int_{-\infty}^{\infty}\rmd x \, \overline{\varphi (x)}
          \frac{1}{\sqrt{2\pi \hbar}} \rme ^{\rmi px/\hbar}
       =\overline{({\cal F}\varphi) (p)} \, .
\end{array}
     \label{definitionketp}
\end{equation}
Note that, although the eigenfunctions 
$\frac{1}{\sqrt{2\pi \hbar}} \rme ^{\rmi px/\hbar}$ are in principle well
defined for any complex $p$, the momentum in Eq.~(\ref{definitionketp}) 
runs only over ${\rm Sp}(P)=(-\infty ,\infty )$, because we are interested 
in assigning
kets $|p\rangle$ only to the momenta in the spectrum of $P$, which are the
only momenta that participate in the Dirac basis expansion associated
to $P$. The eigenfunctions corresponding to the multiplication 
operator are just the delta function $\delta (x-x')$, and therefore the ket 
corresponding to each $x \in {\rm Sp}(Q)$ is defined as
\begin{equation}
\begin{array}{rcl}
       |x\rangle :\mathbf \Phi & \longmapsto & {\mathbb C} \\
       \varphi & \longmapsto & \langle \varphi |x\rangle  := 
       \int_{-\infty}^{\infty}\rmd x' \, \overline{\varphi (x')}
          \delta (x-x')
       =\overline{\varphi (x)} \, .
\end{array}
     \label{definitionketx}
\end{equation}
Similarly, we define the eigenkets corresponding to the Hamiltonian:
\begin{equation}
\hskip-0.7cm
\begin{array}{rcl}
       |E^{\pm}\rangle _{\rm l,r} :\mathbf \Phi & \longmapsto & {\mathbb C} \\
       \varphi & \longmapsto & \langle \varphi |E^{\pm}\rangle _{\rm l,r}  := 
       \int_{-\infty}^{\infty}\rmd x \, \overline{\varphi (x)}
           \chi _{\rm l,r}^{\pm}(x;E) 
       =\overline{(U_{\pm}\varphi) _{\rm l,r}(E)} \, .
\end{array}
    \label{definitionketE}
\end{equation}
Note that in Eq.~(\ref{definitionketE}) we have defined four different 
kets. Note also that, although the eigenfunctions 
$\chi _{\rm l,r}^{\pm}(x;E)$ are in principle well defined for any complex 
$E$, the energy in Eq.~(\ref{definitionketE}) runs only over
${\rm Sp}(H)=[0,\infty )$, because we are interested in assigning kets 
$|E^{\pm}\rangle _{\rm l,r}$ only to the energies in the spectrum of $H$,
which are the only energies that participate in the Dirac basis expansion
associated to $H$.

The following proposition, whose proof can be found in \ref{sec:proofprop},
summarizes the results of this subsection:

\vskip0.5cm

\theoremstyle{plain}
\newtheorem*{Prop1}{Proposition~1}
\begin{Prop1} The triplet of spaces (\ref{RHSCONTpr}) is a 
rigged Hilbert space, and it satisfies all the requirements demanded 
in the Introduction. More specifically,
\begin{itemize}
\item[(i)] The quantities (\ref{nmnorms}) fulfill the conditions to be a norm. 

\item[(ii)] The space $\Sw$ is stable under the 
action of $P$, $Q$ and $H$. The restrictions of $P$, $Q$ and $H$ to
$\Sw$ are essentially self-adjoint, $\tau _{\mathbf \Phi}$-continuous 
operators. The space 
$\Sw$ is dense in $L^2(\mathbb R,\rmd x)$. 

\item[(iii)] The kets $|p\rangle$, $|x\rangle$ and 
$|E^{\pm}\rangle _{\rm l,r}$ are 
well-defined antilinear functionals on $\Sw$, 
i.e., they belong to $\Swt$. 

\item[(iv)] The kets $|p\rangle$ are generalized eigenvectors of $P$,
\begin{equation}
       P|p\rangle=p|p\rangle \, , \quad p\in \mathbb R \, ;
\end{equation}
the kets $|x\rangle$ are generalized eigenvectors of $Q$,
\begin{equation}
       Q|x\rangle=x|x\rangle \, , \quad x \in \mathbb R \, ;
\end{equation}
the kets $|E^{\pm}\rangle _{\rm l,r}$ are generalized eigenvectors of $H$,
\begin{equation}
       H|E^{\pm}\rangle_{\rm l,r} =E|E^{\pm}\rangle _{\rm l,r}
   \, , \quad E\in [0,\infty )  \, .
\end{equation}
\end{itemize}
\end{Prop1}

\vskip0.5cm

(Note that $|p\rangle$ and $|x\rangle$ are in particular tempered 
distributions, whereas $|E^{\pm}\rangle_{\rm l,r}$ are not.)

\subsection{The Dirac bras}

We have constructed the kets $|p\rangle$, $|x\rangle$ and 
$|E^{\pm}\rangle _{\rm l,r}$, and we have shown that they belong to the
space of {\it antilinear} functionals over 
$\Sw$, which we denoted by
$\Swt$. In
this subsection, we construct the corresponding bras $\langle x|$, $\langle p|$
and $_{\rm l,r}\langle ^{\pm}E|$, and we show that they belong to the space
of {\it linear} functionals over $\Sw$, which we 
shall denote by $\Swp$. The triplet of spaces
\begin{equation}
     \rhsSwp
      \label{rhsforbras}
\end{equation}
or, equivalently,
\begin{equation}
       \mathbf \Phi \subset {\cal H}
       \subset 
        \mathbf \Phi ^{\prime}
\end{equation}
is also a rigged Hilbert space, although now suitable to contain the eigenbras
of the observables.

The definition of the bra $\langle p|$ is as follows:
\begin{equation}
\begin{array}{rcl}
       \langle p| :\mathbf \Phi & \longmapsto & {\mathbb C} \\
       \varphi & \longmapsto & \langle p| \varphi \rangle  := 
       \int_{-\infty}^{\infty}\rmd x \, \varphi (x)
          \frac{1}{\sqrt{2\pi \hbar}} \rme ^{-\rmi px/\hbar}
       =({\cal F}\varphi) (p) \, .
\end{array}
     \label{definitionbrap}
\end{equation}
Comparison with Eq.~(\ref{definitionketp}) shows that the action of
$\langle p|$ is the complex conjugate of the action of $|p \rangle$:
\begin{equation}
      \langle p| \varphi \rangle = \overline{\langle \varphi |p \rangle} \, .
\end{equation} 
The bra $\langle x|$ is defined as
\begin{equation}
\begin{array}{rcl}
       \langle x| :\mathbf \Phi & \longmapsto & {\mathbb C} \\
       \varphi & \longmapsto & \langle x| \varphi \rangle  := 
       \int_{-\infty}^{\infty}\rmd x' \, \varphi (x')
          \delta (x-x')
       =\varphi (x) \, .
\end{array}
     \label{definitionbrax}
\end{equation}
Comparison with Eq.~(\ref{definitionketx}) shows that the action of
$\langle x|$ is complex conjugated to the action of $|x \rangle$:
\begin{equation}
      \langle x| \varphi \rangle = \overline{\langle \varphi |x \rangle} \, .
\end{equation} 
Analogously, the eigenbras of the Hamiltonian are defined as
\begin{equation}
   \hskip-0.7cm
\begin{array}{rcl}
       _{\rm l,r}\langle ^{\pm}E| :\mathbf \Phi & \longmapsto & {\mathbb C} \\
       \varphi & \longmapsto & _{\rm l,r}\langle ^{\pm}E|\varphi\rangle := 
       \int_{-\infty}^{\infty}\rmd x \, \varphi (x)
           \overline{\chi _{\rm l,r}^{\pm}(x;E)} 
       =(U_{\pm}\varphi) _{\rm l,r}(E) \, .
\end{array}
    \label{definitionbraE}
\end{equation}
(Note that in Eq.~(\ref{definitionbraE}) we have defined four different 
bras.) Comparison with Eq.~(\ref{definitionketE}) shows that the actions of 
the bras $_{\rm l,r}\langle ^{\pm}E|$ are the complex conjugates of the 
actions of the kets $|E ^{\pm} \rangle _{\rm l,r}$:
\begin{equation}
      _{\rm l,r}\langle ^{\pm}E|\varphi\rangle =      
      \overline{\langle \varphi |E ^{\pm} \rangle}_{\rm l,r} \, .
      \label{braketccE}
\end{equation} 

The bras $\langle p|$, $\langle x|$ and $_{\rm l,r}\langle ^{\pm}E|$ are 
eigenvectors of $P$, $Q$ and $H$, respectively, as the following proposition 
shows:

\vskip0.5cm

\theoremstyle{plain}
\newtheorem*{Prop2}{Proposition~2}
\begin{Prop2} Within the rigged Hilbert space (\ref{rhsforbras}), it holds that
\begin{itemize}

\item[(i)] The bras $\langle p|$, $\langle x|$ and 
$_{\rm l,r} \langle ^{\pm}E|$ are 
well-defined linear functionals over $\Sw$, 
i.e., they belong to $\Swp$. 

\item[(ii)] The bras $\langle p|$ are generalized left-eigenvectors of $P$,
\begin{equation}
       \langle p|P=p\langle p| \, , \quad p\in \mathbb R \, ;
\end{equation}
the bras $\langle x|$ are generalized left-eigenvectors of $Q$,
\begin{equation}
      \langle x|Q=x\langle x| \, , \quad x \in \mathbb R \, ;
\end{equation}
the bras $_{\rm l,r} \langle ^{\pm}E|$ are generalized left-eigenvectors of 
$H$,
\begin{equation}
      _{\rm l,r} \langle ^{\pm}E|H=
       E \hskip0.12cm  
         _{\rm l,r} \langle ^{\pm}E|    \, , \quad E\in [0,\infty )  \, .
\end{equation}
\end{itemize}
\end{Prop2}

\vskip0.5cm

The proof of Proposition~2 is included in \ref{sec:proofprop}.

Note that, in particular, and in accordance with Dirac's prescription, there 
is a one-to-one correspondence between bras and 
kets: Given an observable $A$, to each element $a$ in the spectrum of
$A$, there corresponds a bra $\langle a|$ that is a left-eigenvector of $A$ 
and also a ket $|a\rangle$ that is a right-eigenvector of $A$. The bra 
$\langle a|$ belongs to $\mathbf \Phi ^{\prime}$, whereas the ket $|a\rangle$ 
belongs to $\mathbf \Phi ^{\times}$.

\subsection{The Dirac basis vector expansions}

Another important aspect of Dirac's formalism is that the bras and kets
form a complete basis system such that [see also Eq.~(\ref{diracbve})]
\begin{equation}
  \sum_{\alpha}\int_{{\rm Sp}(A)}  \rmd a \, 
      |a\rangle _{\alpha} \, _{\alpha}\langle a| = I \, .
   \label{resonident}
\end{equation}
In the present subsection, we derive various 
Dirac basis vector expansions for the algebra of the 1D rectangular barrier 
potential.

We start by writing
\begin{equation}
   \langle x|E^{\pm}\rangle _{\rm l,r} := \chi ^{\pm}_{\rm l,r}(x;E) \, , 
\end{equation}
and
\begin{equation}
      _{\rm l,r} \langle ^{\pm}E|x \rangle := 
           \overline{\chi ^{\pm}_{\rm l,r}(x;E)} \, .
\end{equation}
Then, the restriction of (\ref{invdiagonaliza}) to $\Sw$ yields the 
following basis expansions:
\begin{equation}
  \langle x|\varphi \rangle = \int_0^{\infty}\rmd E \,
  \langle x|E^{\pm}\rangle _{\rm l}\, _{\rm l}\langle ^{\pm}E|\varphi \rangle +
  \int_0^{\infty}\rmd E \,
 \langle x|E^{\pm}\rangle _{\rm r}\, _{\rm r}\langle ^{\pm}E|\varphi \rangle 
    \, .
     \label{inveqDva+}
\end{equation}
The restriction of Eqs.~(\ref{hatEl+}) and (\ref{hatEr+}) to $\Sw$ yields four
other basis expansions:
\begin{equation}
      _{\rm l,r}\langle ^{\pm}E|\varphi \rangle =
     \int_{-\infty}^{\infty}\rmd x \
        _{\rm l,r}\langle ^{\pm}E|x\rangle 
      \langle x|\varphi \rangle  \, .
    \label{inveqDvaeix}
\end{equation}

The basis vector expansions (\ref{inveqDva+})-(\ref{inveqDvaeix}) are very 
similar to those given by the restriction of the Fourier transform to 
$\Sw$. If we define
\begin{equation}
      \langle x|p\rangle := \frac{1}{\sqrt{2\pi \hbar }}\rme ^{\rmi px /\hbar}
     \, ,
\end{equation}
\begin{equation}
      \langle p|x\rangle := \frac{1}{\sqrt{2\pi \hbar }}\rme ^{-\rmi px /\hbar}
     \, ,
\end{equation}
then the restriction of (\ref{fourt}) to $\Sw$ 
yields 
\begin{equation}
      \langle p|\varphi \rangle =
     \int_{-\infty}^{\infty}\rmd x \
        \langle p|x\rangle 
      \langle x|\varphi \rangle  \, ,
    \label{inveqDvaeipx}
\end{equation}
whereas the restriction of the inverse of (\ref{fourt}) to 
$\Sw$ yields
\begin{equation}
      \langle x|\varphi \rangle =
     \int_{-\infty}^{\infty}\rmd p \
        \langle x|p\rangle 
      \langle p|\varphi \rangle  \, .
    \label{inveqDvaeixp}
\end{equation}
The similarity between the Dirac basis vector expansions
(\ref{inveqDva+})-(\ref{inveqDvaeix}) and the Fourier
expansions (\ref{inveqDvaeipx})-(\ref{inveqDvaeixp}) is another facet
of the parallel between Dirac's formalism and Fourier methods.

For the sake of completeness, we include the 1D rectangular potential version
of the Nuclear Spectral Theorem~\cite{GELFAND} (see \ref{sec:proofprop} for
its proof):

\vskip0.5cm

\theoremstyle{plain}
\newtheorem*{Prop3}{Proposition~3}
\begin{Prop3} (Nuclear Spectral Theorem) Let 
\begin{equation}
     \rhsSwt
\end{equation}
be the RHS of the 1D rectangular barrier algebra such that $\Sw$
remains invariant under the action of the algebra $\cal A$, and such
that the operators of $\cal A$ are $\tau _{\mathbf \Phi}$-continuous,
essentially self-adjoint 
operators over $\Sw$. Then, for each element in
the spectrum of $P$, $Q$ or $H$, there is a generalized eigenvector such that
\begin{eqnarray}
       &P|p\rangle =p|p\rangle \, , \quad &p \in \mathbb R \, ,
                \label{eigePket} \\
       &Q|x\rangle =x|x\rangle \, , \quad &x \in \mathbb R \, ,
               \label{eigeQket} \\
       &H|E^{\pm}\rangle _{\rm l,r}=E|E^{\pm} \rangle _{\rm l,r} 
                \, , \quad &E\in [0,\infty ) \, ,  \label{eigeHket}
\end{eqnarray}
and such that for all $\varphi ,\psi \in \Sw$
\begin{eqnarray}
      (\varphi ,\psi ) &=& \int_0^{\infty}\rmd E\, 
      \langle \varphi |E^{\pm}\rangle_{\rm l}\, 
       _{\rm l}\langle ^{\pm}E|\psi \rangle +
       \int_0^{\infty}\rmd E\, 
      \langle \varphi |E^{\pm}\rangle_{\rm r}\, 
       _{\rm r}\langle ^{\pm}E|\psi \rangle  \label{spinofEbk}   \\
      &=&\int_{-\infty}^{\infty}\rmd p \, 
      \langle \varphi |p\rangle  \langle p|\psi \rangle  \label{spinofpbk} \\
      &=&\int_{-\infty}^{\infty}\rmd x \, 
      \langle \varphi |x\rangle  \langle x|\psi \rangle  \, ,
       \label{spinofxbk}
\end{eqnarray}
and for all $\varphi ,\psi \in \Sw$, $n=1,2, \ldots$
\begin{equation}
      \hskip-1cm (\varphi ,H^n \psi )= \int_0^{\infty} \rmd E \,
      E^n \langle \varphi |E^{\pm}\rangle_{\rm l}\, 
       _{\rm l}\langle ^{\pm}E|\psi \rangle +
       \int_0^{\infty}\rmd E\, E^n
      \langle \varphi |E^{\pm}\rangle_{\rm r}\, 
       _{\rm r}\langle ^{\pm}E|\psi \rangle  \, , 
        \label{GMT2H}
\end{equation}
\begin{equation}
     (\varphi ,P^n \psi )=\int_{-\infty}^{\infty}\rmd p \, p^n
      \langle \varphi |p\rangle  \langle p|\psi \rangle \, , \label{GMT2P}
\end{equation}
\begin{equation}
     (\varphi ,Q^n \psi )=\int_{-\infty}^{\infty}\rmd x \, x^n 
      \langle \varphi |x\rangle  \langle x|\psi \rangle \, . \label{GMT2Q}
\end{equation}
\end{Prop3}

\vskip0.5cm

If we ``sandwich'' Eq.~(\ref{resonident}) in between
two elements $\varphi$ and $\psi$ of $\Sw$, then
we obtain the expansions (\ref{spinofEbk})-(\ref{GMT2Q}), when 
$A=H^n, P^n, Q^n$, $n=0,1,2,\ldots$. If we ``sandwich'' 
Eq.~(\ref{resonident}) in between an element $\varphi$ of $\Sw$ and a bra
$\langle x|$, $_{\rm l,r}\langle ^{\pm}E|$ or $\langle p|$, then we obtain the
expansions (\ref{inveqDva+})-(\ref{inveqDvaeix}) and 
(\ref{inveqDvaeipx})-(\ref{inveqDvaeixp}). This ``sandwiching,'' however,
is not valid when $\varphi$ or $\psi$ lies outside 
$\Sw$, because then the action of the bras and
kets is not well defined. Thus, the RHS, rather than just the Hilbert space, 
fully justifies Dirac's formalism.

\subsection{Energy, momentum and wave-number representations of the rigged 
Hilbert space}

In subsection~\ref{sec:constr}, we constructed the position representation of 
the rigged Hilbert space of the 1D rectangular barrier algebra 
[see Eq.~(\ref{RHSCONTpr})]. In this subsection, we construct three spectral
representations of (\ref{RHSCONTpr}): the energy, the momentum and the 
wave-number representations. 

We start with the energy representation. The unitary operators $U_{\pm}$ of 
Eq.~(\ref{dirintdec}) afford two energy representations of the RHS 
(\ref{RHSCONTpr}). The energy representations of $\Sw$ will be denoted as
\begin{equation}
       \Swhpm \equiv U_{\pm} \Sw \, .
      \label{spaenrep}
\end{equation}
On $\Swhpm$, the Hamiltonian acts
as the multiplication operator, as Eq.~(\ref{idagoplofh}) shows. The spaces
$\Swhpm$ are linear subspaces
of $L^2([0,\infty ),\rmd E)\oplus L^2([0,\infty ),\rmd E)$. In order to endow 
$\Swhpm$ with a topology, we carry the topology on $\Sw$ onto $\Swhpm$,
\begin{equation}
      \tau _{\widehat{\mathbf \Phi}_{\pm}}:=
      U_{\pm} \tau _{\mathbf \Phi} \, .
\end{equation}
Endowed with these topologies, $\Swhpm$ are linear topological 
spaces. If we denote the antidual spaces of 
$\Swhpm$ by $\Swhpmt$, then we have
\begin{equation}
     U_{\pm}^{\times} \Swt = \left[ U_{\pm} \Sw \right] ^{\times}=
     \Swhpmt  \, .
\end{equation}

We can further split the energy representations of the RHS into left and right 
components, which are associated to left and right incidences. In order to do 
so, we need to recall the definition of the left and right components of the 
wave functions [see Eqs.~(\ref{hatEl+}) and (\ref{hatEr+})]:
\begin{equation}
     \widehat{\varphi}_{\rm l,r}^{\pm}(E)=(U_{\pm}\varphi )_{\rm l,r}(E) \, .
\end{equation}
Any element $\widehat{\varphi}^{\pm}$ of $\Swhpm$ can therefore be written as 
a two-component vector,
\begin{equation}
      \widehat{\varphi}^{\pm} \equiv \left[ \widehat{\varphi}_{\rm l}^{\pm},
      \widehat{\varphi}_{\rm r}^{\pm} \right] \, ,
\end{equation}
which is equivalent to write the spaces (\ref{spaenrep}) as sums of left
and right components:
\begin{equation}
      \Swhpm \equiv \Swhpml \oplus \Swhpmr  \, .
      \label{splishat} 
\end{equation}
Their antiduals can be split in a similar way,
\begin{equation}
    \Swhpmt \equiv \Swhpmlt \oplus \Swhpmrt  \, .
       \label{splishatandual}  
\end{equation}
The energy representation of the kets $|E^{\pm}\rangle _{\rm l,r}$ is 
given by a familiar distribution. If we denote the energy representation of 
these kets by $|\widehat{E}^{\pm}\rangle _{\rm l,r}$, then the following 
equalities
\begin{eqnarray}
    \langle \widehat{\varphi}_{\rm l,r}^{\pm}|
      \widehat{E}^{\pm}\rangle _{\rm l,r} &=&
   \langle \widehat{\varphi}_{\rm l,r}^{\pm}|
        U_{\pm}^{\times}|E^{\pm}\rangle _{\rm l,r} \\ &=&
    \langle U_{\pm}^{-1}\widehat{\varphi}_{\rm l,r}^{\pm}|
           E^{\pm}\rangle_{\rm l,r}  \nonumber \\
     &=& \int_{-\infty}^{\infty}\rmd x \, 
         \overline{\varphi (x)} \chi_{\rm l,r}^{\pm} (x;E)
          \nonumber \\
       &=& \overline{\widehat{\varphi}_{\rm l,r}^{\pm}(E)} \, 
\end{eqnarray}
show that $|\widehat{E}^{\pm}\rangle _{\rm l}$ and 
$|\widehat{E}^{\pm}\rangle _{\rm r}$ act as the {\it antilinear}
Schwartz delta functional over the spaces $\Swhpml$ and $\Swhpmr$, 
respectively.

The different realizations of the RHS are easily visualized through the 
following diagram:
\begin{equation}
     \hskip-2.5cm \begin{array}{cccccccccc}
      H; \ \varphi &  & \Sw &
      \subset & L^2(\mathbb R, \rmd x)  &
 \subset &  \Swt &
     & |E^{\pm}\rangle _{\rm l,r}   \nonumber \\  
       &   & \downarrow U_{\pm}  &  &
        \downarrow U_{\pm}      &
       & \downarrow U_{\pm}^{\times} & 
         &   \nonumber \\   
      \widehat{H}; \ \widehat{\varphi}_{\pm} &  &
        \Swhpm  & 
       \subset & 
     \oplus  L^2([0,\infty ), \rmd E)  & \subset & 
        \Swhpmt & 
      &  |\widehat{E}^{\pm}\rangle _{\rm l,r}  \\ [1ex]
      \end{array}
      \label{diagramsavp}
\end{equation}
where $\oplus  L^2([0,\infty ), \rmd E)$ denotes
$L^2([0,\infty ), \rmd E)\oplus  L^2([0,\infty ), \rmd E)$. The top line 
of diagram~(\ref{diagramsavp}) displays the Hamiltonian, the wave 
functions, the RHS and the Dirac kets in the position representation. The 
bottom line displays their energy representation counterparts.

We can also construct the energy representation of the eigenbras
$_{\rm l,r} \langle ^{\pm}E|$. To this end, we first construct the energy 
representation of $\Swp$, which we 
shall denote by $\Swhpmp$. These two spaces are related as follows:
\begin{equation}
     U_{\pm}^{\prime} \Swp =
     \left[ U_{\pm} \Sw \right] ^{\prime}=
     \Swhpmp \, .
\end{equation}
Similarly to Eqs.~(\ref{splishat}) and (\ref{splishatandual}), the dual space
can be split into left and right components,
\begin{equation}
    \Swhpmp =\Swhpmlp \oplus \Swhpmrp \, .
\end{equation}
Now, if we denote the energy representation of the energy eigenbras
by $_{\rm l,r} \langle ^{\pm}\widehat{E}|$, then the following 
equalities
\begin{eqnarray}
      _{\rm l,r} \langle ^{\pm}\widehat{E}|
          \widehat{\varphi}_{\rm l,r}^{\pm} \rangle  &=&
      _{\rm l,r} \langle ^{\pm}E|U_{\pm}^{\prime}|
          \widehat{\varphi}_{\rm l,r}^{\pm} \rangle  \\
       &=& _{\rm l,r} \langle ^{\pm}E|U_{\pm}^{-1}
          \widehat{\varphi}_{\rm l,r}^{\pm} \rangle  \nonumber \\
     &=& \int_{-\infty}^{\infty}\rmd x \, 
         \varphi (x) \overline{\chi_{\rm l,r}^{\pm} (x;E)}
          \nonumber \\
       &=& \widehat{\varphi}_{\rm l,r}^{\pm}(E) 
\end{eqnarray}
show that $_{\rm l} \langle ^{\pm}\widehat{E}|$ and 
$_{\rm r} \langle ^{\pm}\widehat{E}|$ are the {\it linear}
Schwartz delta functional over the spaces $\Swhpml$ and $\Swhpmr$,
respectively. 

The diagram corresponding to the bras is as follows:
\begin{equation}
     \hskip-2cm \begin{array}{cccccccccc}
      H; \ \varphi &  & \Sw &
      \subset & L^2(\mathbb R, \rmd x)  &
 \subset & \Swp &
     &  _{\rm l,r} \langle ^{\pm}E|   \nonumber \\  
       &  & \downarrow U_{\pm}  &  &
        \downarrow U_{\pm}      &
       & \downarrow U_{\pm}^{\prime} & 
         &   \nonumber \\   
      \widehat{H}; \ \widehat{\varphi}_{\pm} &  &
         \Swhpm & \subset & 
     \oplus  L^2([0,\infty ), \rmd E)  & \subset & 
        \Swhpmp & 
      &  _{\rm l,r} \langle ^{\pm} \widehat{E}|  \\ [1ex]
      \end{array}
      \label{diagramsavpbra}
\end{equation}
The energy representations of $P$ and $Q$ have not been included, since they
are fairly complicated.

The momentum representation of the RHS (\ref{RHSCONTpr}) can be constructed
in a similar fashion, by way of the Fourier transform ${\cal F}$. We shall not
reproduce the calculations here but only provide the resulting 
diagrams. The diagram corresponding to the position and 
momentum kets reads as
\begin{equation}
     \hskip-1cm \begin{array}{cccccccccc}
      P,Q; \ \varphi &  & \Sw &  \subset & L^2(\mathbb R, \rmd x)  &
 \subset & \Swt &  & |p\rangle , \, |x\rangle    \nonumber \\  
       &   & \downarrow {\cal F}  &  &
        \downarrow {\cal F}      &
       & \downarrow {\cal F}^{\times} & 
         &   \nonumber \\   
      \widehat{P},\widehat{Q}; \ \widehat{\varphi} &  &
         \Swh & 
       \subset & L^2(\mathbb R , \rmd p)  & \subset & 
        \Swht & 
      &  |\widehat{p}\rangle , \, |\widehat{x}\rangle  \\ [1ex]
      \end{array}
      \label{diagramsavpFk}
\end{equation}
where $\widehat{P}$ acts as the multiplication operator by $p$, $\widehat{Q}$
acts as the differential operator $\rmi \hbar \rmd / \rmd p$,
$|\widehat{p}\rangle$ is the {\it antilinear} Schwartz delta functional, and
$|\widehat{x}\rangle$ is the {\it antilinear} functional whose kernel is 
$(2\pi \hbar)^{-1/2}  \exp \left(-\rmi px/\hbar\right)$. The momentum diagram 
for the position and momentum bras is
\begin{equation}
     \hskip-1cm \begin{array}{cccccccccc}
      P,Q; \ \varphi &  & \Sw &
      \subset & L^2(\mathbb R, \rmd x)  &
 \subset & \Swp &
     & \langle p| , \, \langle x|    \nonumber \\  
       &   & \downarrow {\cal F}  &  &
        \downarrow {\cal F}      &
       & \downarrow {\cal F}^{\prime} & 
         &   \nonumber \\   
      \widehat{P},\widehat{Q}; \ \widehat{\varphi} &  &
         \Swh& 
       \subset & L^2(\mathbb R , \rmd p)  & \subset & 
        \Swhp & 
      &  \langle\widehat{p}| , \, \langle \widehat{x}|  \\ [1ex]
      \end{array}
      \label{diagramsavpFb}
\end{equation}
where $\langle\widehat{p}|$ is the {\it linear} Schwartz delta functional, and
$\langle \widehat{x}|$ is the {\it linear} functional with kernel 
$(2\pi \hbar) ^{-1/2} \exp \left(\rmi px/\hbar\right)$. The momentum 
representation of $H$ has not been included, since in the momentum 
representation $H$ has a complicated expression. 

The momentum representation should not be confused with the wave number
representation, which we construct in the remainder of this subsection.

The eigenfunctions of the Schr\"odinger differential operator, the Green 
function and the transmission and reflection coefficients depend on
the square root of the energy rather than on the energy itself. Thus,
the wave number, which is defined as
\begin{equation}
        k:=\sqrt{\frac{2m}{\hbar ^2} \, E} \, , 
       \label{momenuks}
\end{equation}
is a more convenient variable. In terms of $k$, the $\delta$-normalized 
eigensolutions of the differential operator (\ref{doh}) read as
\begin{equation}
      \langle x|k^{\pm}\rangle _{\rm l,r} \equiv
       \sqrt{\frac{\hbar ^2k}{m}\ } \, \chi _{\rm l,r}^{\pm}(x;E) \, .
      \label{continuoseign}
\end{equation}
These eigensolutions can be used to obtain the unitary operators 
$V_{\pm}$ that transform between the position and the wave-number 
representations,
\begin{equation}  
     \hskip-1.2cm  \widehat{\widehat{f}\,}_{\pm}(k)=(V_{\pm}f)(k)=
      \int_{-\infty}^{\infty}\rmd x \, f(x) 
          \overline{\langle x|k^{\pm}\rangle} _{\rm l} +
         \int_{-\infty}^{\infty}\rmd x \, f(x) 
          \overline{\langle x|k^{\pm}\rangle} _{\rm r}   \, ,
       \ f\in {\cal H} \, ,
    \label{Vcontinuoseign}
\end{equation}
where ``$\widehat{\widehat{\quad}}$'' denotes the 
$k$-representation. On this representation, the Hamiltonian acts as 
multiplication by $\frac{\hbar ^2}{2m}k^2$. To each $k\in [0,\infty )$, there 
correspond four eigenkets $|k^{\pm}\rangle_{\rm l,r}$ that act on $\Sw$ as 
the following integral operators:  
\begin{equation}
      \hskip-0.5cm  \langle \varphi |k^{\pm}\rangle_{\rm l,r} :=
       \int_{-\infty}^{\infty}\rmd x \,
       \langle \varphi |x\rangle 
      \langle x|k^{\pm}\rangle _{\rm l,r} =
       \overline{(V_{\pm}\varphi )_{\rm l,r} (k)}\, , \quad
        \varphi \in \Sw  \, . 
\end{equation}
These eigenkets are generalized eigenvectors of the Hamiltonian with
eigenvalue $\frac{\hbar ^2}{2m}k^2$. The following diagram provides the
$k$-representation counterpart of~(\ref{diagramsavp}):
\begin{equation}
     \hskip-2.5cm \begin{array}{cccccccccc}
      H; \ \varphi &  & \Sw &
      \subset & L^2(\mathbb R, \rmd x)  &
 \subset & \Swt &
     & |k^{\pm}\rangle _{\rm l,r}   \nonumber \\  
       &   & \downarrow V_{\pm}  &  &
        \downarrow V_{\pm}      &
       & \downarrow V_{\pm}^{\times} & 
         &   \nonumber \\   
      \widehat{\widehat{H}}; \ \widehat{\widehat{\varphi}\,}_{\pm} &  &
         \Swhhpm& 
       \subset & 
     \oplus  L^2([0,\infty ), \rmd k)  & \subset & 
      \Swhhpmt
     &  &  |\widehat{\widehat{k}}\,^{\pm}\rangle _{\rm l,r}  
                                              \\ [1ex]
      \end{array}
      \label{kdiagramsavp}
\end{equation}
where $\oplus  L^2([0,\infty ), \rmd k)$ denotes
$L^2([0,\infty ), \rmd k) \oplus  L^2([0,\infty ), \rmd k)$, and
$|\widehat{\widehat{k}} \, ^{\pm}\rangle _{\rm l,r}$ act as the 
{\it antilinear} Schwartz delta functional. The $k$-representation
counterpart of~(\ref{diagramsavpbra}) is given by the following diagram:
\begin{equation}
     \hskip-2.5cm \begin{array}{cccccccccc}
      H; \ \varphi &  & \Sw  &
      \subset & L^2(\mathbb R, \rmd x)  & \subset &  \Swp &
     & _{\rm l,r}\langle ^{\pm} k|   \nonumber \\  
       &   & \downarrow V_{\pm}  &  &
        \downarrow V_{\pm}      &
       & \downarrow V_{\pm}^{\prime} & 
         &   \nonumber \\   
      \widehat{\widehat{H}}; \ \widehat{\widehat{\varphi}\,}\!_{\pm} &  &
         \Swhhpm & 
       \subset & 
     \oplus  L^2([0,\infty ), \rmd k)  & \subset & \Swhhpmp & 
      & _{\rm l,r}\langle ^{\pm}\widehat{\widehat{k}}| \\ [1ex]
      \end{array}
      \label{kdiagramsavpbra}
\end{equation}
where $_{\rm l,r}\langle ^{\pm}\widehat{\widehat{k}}|$ act as the 
{\it linear} Schwartz delta functional.

The wave number is particularly useful in writing the Green function in 
a simple, compact form, as we 
are going to see now. Expressions (\ref{tk}) for $T(k)$ and 
(\ref{tstark}) for $T^*(k)$ yield
\begin{equation}
       T(-k)=T^*(k) \, , \quad k>0 \, . 
       \label{relttstar}
\end{equation}
From Eqs.~(\ref{chir+}) and (\ref{chir-}) it results that
\begin{equation}
      \chi _{\rm r}^+(x;-k) = \rmi  \chi _{\rm r}^-(x;k) \, , \quad k>0 \, ,
\end{equation}
and from Eqs.~(\ref{chil+}) and (\ref{chil-}) it results that
\begin{equation}
      \chi _{\rm l}^+(x;-k) = \rmi \chi _{\rm l}^-(x;k) \, , \quad k>0 \, .
\end{equation}
We can use the last three equations to write the Green function for all values 
of $k$ (and therefore for all values of $E$):
\begin{equation}
      G(x,x';k)=\frac{2\pi}{\rmi} \,
         \frac{\chi _{\rm r}^+(x_<;k)\chi _{\rm l}^+(x_>;k)}{T(k)} \, , 
        \quad k\in {\mathbb C} \, ,  
      \label{grenkfunc}
\end{equation}
where $x_< , x_>$ refer to the smaller and to the bigger of $x$ and $x'$,
respectively.

\section{Conclusions}
\label{sec:conclusions}

We have explicitly constructed the RHSs of the algebra of the 1D rectangular
potential. In the position representation, these RHSs are given by
\begin{equation}
     \Sw \subset L^2(\mathbb{R},\rmd x) \subset  \Swt  \, , 
\end{equation}
\begin{equation}
     \Sw \subset L^2(\mathbb{R},\rmd x) \subset \Swp \, . 
\end{equation}
On $\Sw$, the observables are essentially self-adjoint, continuous 
operators. Algebraic operations such as commutation relations are
well defined on $\Sw$.

We have also constructed the Dirac bras and kets of each observable of the
algebra, as well as the basis expansions generated by the bras and kets. By
doing so, we have shown (once again) that the RHS fully accounts for 
Dirac's formalism.

By comparing the results for the Fourier transform $\cal F$ with those for
the unitary operators $U_{\pm}$,
we have seen that Dirac's formalism can be viewed as an extension of
Fourier methods: Monoenergetic eigenfunctions extend the notion of
monochromatic plane waves, $U_{\pm}$ extend the notion of Fourier transform, 
and Dirac's basis expansions extend the notion of Fourier decomposition.

The results of this paper can be applied to many other algebras, at least
when resonance eigenvalues are not considered. In general, the space of test 
functions $\mathbf \Phi$ is given by the maximal invariant subspace of the 
algebra, and the spaces of distributions $\mathbf \Phi ^{\times}$ and 
$\mathbf \Phi ^{\prime}$ are given by the antidual and dual spaces of 
$\mathbf \Phi$. 

As a corollary to the results of this paper, we can derive the RHSs of the 
algebra of the 1D free Hamiltonian. By making $V_0$ tend to zero, we can see
that these RHSs are given by 
\begin{equation}
     \mathcal{S}(\mathbb{R}) \subset L^2(\mathbb{R},\rmd x) \subset
     \mathcal{S}^{\times}(\mathbb{R}) \, , 
\end{equation}
\begin{equation}
     \mathcal{S}(\mathbb{R}) \subset L^2(\mathbb{R},\rmd x) \subset
     \mathcal{S}^{\prime}(\mathbb{R}) \, , 
\end{equation}
where $\mathcal{S}(\mathbb{R})$ is the Schwartz space.

Finally, of mathematical interest is the introduction of a new space of
test functions, the Schwartz-like space $\Sw$, and new spaces of 
distributions, the spaces of tempered-like distributions $\Swt$ and $\Swp$.

\ack

Research supported by the Basque Government through reintegration 
fellowship No.~BCI03.96.

\appendix
\setcounter{section}{0}

\section{Auxiliary functions}
\label{sec:appauxfunc}

For the sake of completeness, we provide the explicit expressions of the
coefficients of the eigenfunctions:

\begin{eqnarray}
     &\widetilde{T}(\widetilde{k})=\rme ^{\widetilde{k}(b-a)} 
   \frac{ 4\widetilde{Q}/\widetilde{k}}
      {(1+\widetilde{Q}/\widetilde{k})^2 \rme ^{\widetilde{Q}(b-a)}-
      (1-\widetilde{Q}/\widetilde{k})^2 \rme ^{-\widetilde{Q}(b-a)}} \\ [1ex]
   &\widetilde{A}_{\rm r}(\widetilde{k})=\frac{-2\rme ^{\widetilde{k}b}
     \rme ^{\widetilde{Q}a} (1-\widetilde{Q}/\widetilde{k})}
        {(1+\widetilde{Q}/\widetilde{k})^2 \rme ^{\widetilde{Q}(b-a)}-
      (1-\widetilde{Q}/\widetilde{k})^2 \rme ^{-\widetilde{Q}(b-a)}} \\ [1ex]
   &\widetilde{B}_{\rm r}(\widetilde{k})=\frac{2\rme ^{\widetilde{k}b}
          \rme ^{-\widetilde{Q}a} (1+\widetilde{Q}/\widetilde{k})}
        {(1+\widetilde{Q}/\widetilde{k})^2 \rme ^{\widetilde{Q}(b-a)}-
      (1-\widetilde{Q}/\widetilde{k})^2 \rme ^{-\widetilde{Q}(b-a)}} \\ [1ex]
   &\widetilde{R}_{\rm r}(\widetilde{k})=\rme ^{2\widetilde{k}b} 
        \frac{(1-(\widetilde{Q}/\widetilde{k})^2 ) 
         \rme ^{\widetilde{Q}(b-a)}-( 1-(\widetilde{Q}/\widetilde{k})^2 ) 
         \rme ^{-\widetilde{Q}(b-a)}}
     {(1+\widetilde{Q}/\widetilde{k})^2 \rme ^{\widetilde{Q}(b-a)}-
      (1-\widetilde{Q}/\widetilde{k})^2 \rme ^{-\widetilde{Q}(b-a)}} \\ [1ex] 
    &\widetilde{R}_{\rm l}(\widetilde{k})=\rme ^{-2\widetilde{k}a} 
        \frac{( 1-(\widetilde{Q}/\widetilde{k})^2 ) 
         \rme ^{\widetilde{Q}(b-a)}
               - ( 1-(\widetilde{Q}/\widetilde{k})^2 ) 
         \rme ^{-\widetilde{Q}(b-a)}}
     {(1+\widetilde{Q}/\widetilde{k})^2 \rme ^{\widetilde{Q}(b-a)}-
      (1-\widetilde{Q}/\widetilde{k})^2 \rme ^{-\widetilde{Q}(b-a)}} \\ [1ex]
    &\widetilde{A}_{\rm l}(\widetilde{k})=\frac{2\rme ^{-\widetilde{k}a} 
         \rme ^{\widetilde{Q}b}(1+\widetilde{Q}/\widetilde{k})}
     {(1+\widetilde{Q}/\widetilde{k})^2 \rme ^{\widetilde{Q}(b-a)}-
      (1-\widetilde{Q}/\widetilde{k})^2 \rme ^{-\widetilde{Q}(b-a)}} \\ [1ex]
   &\widetilde{B}_{\rm l}(\widetilde{k})=\frac{-2\rme ^{-\widetilde{k}a} 
         \rme ^{-\widetilde{Q}b}(1-\widetilde{Q}/\widetilde{k})}
     {(1+\widetilde{Q}/\widetilde{k})^2 \rme ^{\widetilde{Q}(b-a)}-
      (1-\widetilde{Q}/\widetilde{k})^2 \rme ^{-\widetilde{Q}(b-a)} } 
\end{eqnarray}     

\begin{eqnarray}
      &T(k)=\rme ^{-\rmi k(b-a)} \frac{-4Q/k}
                    {(1-Q/k)^2 \rme ^{\rmi Q(b-a)}-
                      (1+Q/k)^2 \rme ^{-\rmi Q(b-a)}}
                \label{tk} \\ [1ex]
     &A_{\rm r}(k)=\frac{2\rme ^{-\rmi kb} \rme ^{-\rmi Qa}(1-Q/k)}
     {(1-Q/k)^2 \rme ^{\rmi Q(b-a)}-(1+Q/k)^2 \rme ^{-\rmi Q(b-a)} } \\ [1ex]
     &B_{\rm r}(k)=\frac{-2\rme ^{-\rmi kb} \rme ^{\rmi Qa}(1+Q/k)}
                    {(1-Q/k)^2 \rme ^{\rmi Q(b-a)}-
                      (1+Q/k)^2 \rme ^{-\rmi Q(b-a)}} \\ [1ex]
     &R_{\rm r}(k)=\rme ^{-2\rmi kb} \frac{
                  \left( 1-(Q/k)^2 \right) \rme ^{\rmi Q(b-a)}
               -\left( 1-(Q/k)^2 \right) \rme ^{-\rmi Q(b-a)}}
                    {(1-Q/k)^2 \rme ^{\rmi Q(b-a)}-
                      (1+Q/k)^2 \rme ^{-\rmi Q(b-a)}} \\ [1ex]
      &R_{\rm l}(k)=\rme ^{2\rmi ka} \frac{
                  \left( 1-(Q/k)^2 \right) \rme ^{\rmi Q(b-a)}
               -\left( 1-(Q/k)^2 \right) \rme ^{-\rmi Q(b-a)}}
                    {(1-Q/k)^2 \rme ^{\rmi Q(b-a)}-
                      (1+Q/k)^2 \rme ^{-\rmi Q(b-a)}} \\ [1ex]
     &A_{\rm l}(k)=\frac{-2\rme ^{\rmi ka} \rme ^{-\rmi Qb}(1+Q/k)}
                    {(1-Q/k)^2 \rme ^{\rmi Q(b-a)}-
                      (1+Q/k)^2 \rme ^{-\rmi Q(b-a)}} \\ [1ex]
     &B_{\rm l}(k)=\frac{2\rme ^{\rmi ka} \rme ^{\rmi Qb}(1-Q/k)}
     {(1-Q/k)^2 \rme ^{\rmi Q(b-a)}-(1+Q/k)^2 \rme ^{-\rmi Q(b-a)} } 
\end{eqnarray}     

\begin{eqnarray}
    &T^*(k)=\rme ^{\rmi k(b-a)} \frac{-4Q/k}
                    {(1-Q/k)^2 \rme ^{-\rmi Q(b-a)}-
                      (1+Q/k)^2 \rme ^{\rmi Q(b-a)}}
          \label{tstark} \\ [1ex]
    &A_{\rm r}^*(k)=\frac{2\rme ^{\rmi kb} \rme ^{\rmi Qa}(1-Q/k)}
     {(1-Q/k)^2 \rme ^{-\rmi Q(b-a)}-(1+Q/k)^2 \rme ^{\rmi Q(b-a)} } \\ [1ex] 
    &B_{\rm r}^*(k)=\frac{-2\rme ^{\rmi kb} \rme ^{-\rmi Qa}(1+Q/k)}
                    {(1-Q/k)^2 \rme ^{-\rmi Q(b-a)}-
                      (1+Q/k)^2 \rme ^{\rmi Q(b-a)}} \\ [1ex]
   &R_{\rm r}^*(k)=\rme ^{2\rmi kb} \frac{
                  \left( 1-(Q/k)^2 \right) \rme ^{-\rmi Q(b-a)}
               -\left( 1-(Q/k)^2 \right) \rme ^{\rmi Q(b-a)}}
                    {(1-Q/k)^2 \rme ^{-\rmi Q(b-a)}-
                      (1+Q/k)^2 \rme ^{\rmi Q(b-a)}}  \\ [1ex]
    &R_{\rm l}^*(k)=\rme ^{-2\rmi ka} \frac{
                  \left( 1-(Q/k)^2 \right) \rme ^{-\rmi Q(b-a)}
               -\left( 1-(Q/k)^2 \right) \rme ^{\rmi Q(b-a)}}
                    {(1-Q/k)^2 \rme ^{-\rmi Q(b-a)}-
                      (1+Q/k)^2 \rme ^{\rmi Q(b-a)}}  \\ [1ex] 
    &A_{\rm l}^*(k)=\frac{-2\rme ^{-\rmi ka} \rme ^{\rmi Qb}(1+Q/k)}
                    {(1-Q/k)^2 \rme ^{-\rmi Q(b-a)}-
                      (1+Q/k)^2 \rme ^{\rmi Q(b-a)}} \\ [1ex]
    &B_{\rm l}^*(k)=\frac{2\rme ^{-\rmi ka} \rme ^{-\rmi Qb}(1-Q/k)}
     {(1-Q/k)^2 \rme ^{-\rmi Q(b-a)}-(1+Q/k)^2 \rme ^{\rmi Q(b-a)} }
\end{eqnarray}

\section{Spectral measures associated to 
$\{ \chi _{\rm l}^-, \chi _{\rm r}^-\}$}
\label{sec:compfor-}

The ``final'' basis $\{ \chi _{\rm l}^-,\chi _{\rm r}^-\}$ can be used as well 
as the ``initial'' basis $\{ \chi _{\rm l}^+,\chi _{\rm r}^+\}$ to calculate 
${\rm Sp}(H)$. This calculation, which follows the procedure of 
Sec.~\ref{sec:spectrum}, is provided in this appendix.

If we choose
\begin{eqnarray}
      &&\sigma _1(x;E)=\chi _{\rm l}^-(x;E)
      \label{sigma1=sci-} \\ 
      &&\sigma _2(x;E)= \chi _{\rm r}^-(x;E)
       \label{sigam2cos-}  
\end{eqnarray}
as the basis of Theorem~3,
then Eqs.~(\ref{chil+}), (\ref{chir-}), (\ref{chil-}), 
(\ref{comcomr+r-}), (\ref{sigma1=sci-}) and (\ref{sigam2cos-}) lead to
\begin{eqnarray}
    &&\chi _{\rm l}^+(x;E)=-\frac{T(E)R_{\rm r}^*(E)}{T^*(E)}\sigma _1(x;E)+
       T(E)\sigma _2(x;E)\, ,
      \label{chir+s1s2-} \\
      &&\chi _{\rm r}^-(x';E)=T^*(E)\overline{\sigma _1(x';\overline{E})}
    -\frac{R_{\rm l}(E)T^*(E)}{T(E)}\overline{\sigma _2(x';\overline{E})} \, . 
      \label{chil-s1s2-} 
\end{eqnarray}
By substituting Eq.~(\ref{chir+s1s2-}) into Eq.~(\ref{green++}) and after
some calculations, we get to
\begin{eqnarray}
     \hskip-0.6cm G(x,x';E)=
       \frac{2\pi}{\rmi} \,
       \left[-\frac{R_{\rm r}^*(E)}{T^*(E)}
    \sigma _1 (x;E) \overline{\sigma _2(x';\overline{E})}+
     \sigma _2(x;E) \overline{\sigma _2(x';\overline{E})} \right]  
        \, , \nonumber \\
      \qquad \hskip5.2cm  \mbox{Re}(E)>0, \mbox{Im}(E)>0\, , \, x>x' \, . 
        \label{redaot++-}
\end{eqnarray}
By substituting Eq.~(\ref{chil-s1s2-}) into Eq.~(\ref{green+-}) and after
some calculations, we get to
\begin{eqnarray}
      && \hskip-0.4cm G(x,x';E)=
       \frac{2\pi}{\rmi} \,
       \left[ -\sigma _1(x;E)
         \overline{\sigma _1 (x';\overline{E})}
     +\frac{R_{\rm l}(E)}{T(E)} 
     \sigma _1(x;E) \overline{\sigma _2 (x';\overline{E})} \right]      
       \, , \nonumber \\
      &&\qquad \hskip5.3cm \mbox{Re}(E)>0, \mbox{Im}(E)<0\, , \, x>x' \, .
      \label{redaot+--}
\end{eqnarray} 
By comparing (\ref{greenfunitthes}) to (\ref{redaot++-}) we obtain
\begin{equation}
      \theta _{ij}^+(E)= \left(  \begin{array}{cc}
       0 & -\frac{2\pi}{\rmi} \frac{R_{\rm r}^*(E)}{T^*(E)}\\
       0 & \frac{2\pi}{\rmi}
                                \end{array}
        \right) 
       \, , \quad  \mbox{Re}(E)>0 \, , \  \mbox{Im}(E)>0 \, .
       \label{theta++-}
\end{equation}
By comparing (\ref{greenfunitthes}) to (\ref{redaot+--}) we obtain
\begin{equation}
      \theta _{ij}^+(E)= \left(  \begin{array}{cc}
      -\frac{2\pi}{\rmi}  & \frac{2\pi}{\rmi} 
                                  \frac{R_{\rm l}(E)}{T(E)} \\
                             0  & 0
                                \end{array}
        \right) 
       \, , \quad  \mbox{Re}(E)>0 \, , \  \mbox{Im}(E)<0 \, .
      \label{theta+--}
\end{equation}
As expected, the functions $\theta _{11}^+(E)$ and $\theta _{22}^+(E)$ both 
have a branch cut along the spectrum of $H$. 

The measures $\varrho _{ij}$ of Theorem~3 can be readily obtained from 
Eqs.~(\ref{theta++-}) and (\ref{theta+--}). The measure $\varrho _{21}$ is 
clearly zero. So is the measure $\varrho _{12}$, since
\begin{equation}
       \varrho _{12}((E_1,E_2)) =\int_{E_1}^{E_2} 
       - \left( \frac{R_{\rm l}(E)}{T(E)}-\frac{R_{\rm r}^*(E)}{T^*(E)}\right) 
           \rmd E = 0  \, .
\end{equation}
The measures $\varrho _{11}$ and $\varrho _{22}$ are simply the
Lebesgue measure:
\begin{equation}
       \varrho _{11}((E_1,E_2))=\varrho _{22}((E_1,E_2))
      =\int_{E_1}^{E_2} \rmd E = E_2-E_1 \, .
\end{equation}

\section{Proofs of propositions}
\label{sec:proofprop}

In this appendix, we provide the proofs of some propositions we stated
in the main body of the paper. For the sake of clarity, in the proofs we 
shall denote the antidual and dual extensions of $H$ by respectively 
$H^{\times}$ and $H^{\prime}$.

\begin{proof}[Proof of Proposition~1] ({\it i}) This is immediate.

\vskip0.3cm

({\it ii}) From the definition of $\cal D$, Eq.~(\ref{ddomain}), and from
the expressions of the differential operators associated to $P$, $Q$ and 
$H$, it can be seen after straightforward (though tedious) calculations
that $\cal D$ is stable under the algebra of observables. It is also easy
to see that $\cal D$ is indeed the largest subdomain of 
$L^2(\mathbb R,\rmd x)$ that remains stable under the action of the algebra
of observables, i.e., $\cal D$ is the maximal invariant subspace of $\cal A$.

That $P$, $Q$ and $H$ are essentially self-adjoint over $\Sw$ is obvious, since
their only possible self-adjoint extensions are those with domains 
${\cal D}(P)$, ${\cal D}(Q)$ and ${\cal D}(H)$.

In order to prove that $H$ is 
$\tau _{\mathbf \Phi}$-continuous, we just have to realize that
\begin{eqnarray}
      \| H \varphi \| _{n,m,l}&=&
     \| P^nQ^mH^lH\varphi \|  \nonumber \\
       &= & \| \varphi \| _{n,m,l+1} \, .
       \label{tauphiscont}
\end{eqnarray}

In order to prove that $P$ and $Q$ are $\tau _{\mathbf \Phi}$-continuous, we 
need the following commutation relations:
\begin{equation}
    [H^n,Q]= -\frac{1}{\rm m}n\rmi \hbar PH^{n-1} \, , \ n=1,2, \ldots
        \label{crHnQ}
\end{equation}
where $\rm m$ refers to the mass,
\begin{equation}
      [Q^n,P]=n\rmi \hbar Q^{n-1} \, , \ n=1,2, \ldots 
        \label{crQnP}
\end{equation}
and
\begin{equation}
      [H^n,P]=0 \, , \ n=1,2, \ldots . 
        \label{crHnP}
\end{equation}
(Note that all these commutation relations are well defined on 
$\mathbf \Phi$.) Then, the $\tau _{\mathbf \Phi}$-continuity of $P$ follows 
from 
\begin{eqnarray}
      \| P \varphi \| _{n,m,l}&=&
     \| P^nQ^mH^lP\varphi \|  \nonumber \\
       &=& \| P^nQ^mPH^l\varphi \| \hskip1cm \mbox{from (\ref{crHnP})} 
                                                    \nonumber \\
       &=& \| P^n(PQ^m+m\rmi \hbar Q^{m-1})H^l\varphi \|
           \hskip1cm \mbox{from (\ref{crQnP})} \nonumber \\
      &\leq & \| P^{n+1}Q^mH^l\varphi \| + 
            m\hbar \| P^{n}Q^{m-1}H^l\varphi \| \nonumber \\
      &=& \| \varphi \|_{n+1,m,l} +m \hbar \| \varphi \|_{n,m-1,l} \, ,
\end{eqnarray}
and the $\tau _{\mathbf \Phi}$-continuity of $Q$ follows from 
\begin{eqnarray}
     \hskip-1cm  \| Q \varphi \| _{n,m,l}&=&
     \| P^nQ^mH^lQ\varphi \|  \nonumber \\
       &=& \| P^nQ^m (QH^l-\frac{1}{\rm m} l\rmi \hbar PH^{l-1})\varphi \|
      \hskip1cm  \mbox{from (\ref{crHnQ})}  \nonumber \\
       &\leq & \| P^nQ^{m+1}H^l\varphi \| + \frac{1}{\rm m}l\hbar 
         \| P^nQ^mPH^{l-1}\varphi \|
           \nonumber \\
       &=& \| \varphi \| _{n,m+1,l} + \frac{1}{\rm m}l\hbar
           \| P^n(PQ^m+m\rmi \hbar Q^{m-1})H^{l-1}\varphi \|
        \hskip1cm  \mbox{from (\ref{crQnP})}  \nonumber \\
       &\leq & \| \varphi \| _{n,m+1,l} + \frac{1}{\rm m}l\hbar
         \left( \| P^{n+1}Q^mH^{l-1}\varphi \| +
             m\hbar \| P^nQ^{m-1}H^{l-1}\varphi \| \right) \nonumber \\
       &=& \| \varphi \| _{n,m+1,l} + \frac{1}{\rm m}l\hbar
          \| \varphi \|_{n+1,m,l-1} +
           \frac{1}{\rm m}l\hbar ^2 m \| \varphi \|_{n,m-1,l-1}  \, .
\end{eqnarray}

In order to show that $\Sw$ is dense in 
$L^2(\mathbb R, \rmd x)$, we need to define the space of infinitely
differentiable functions with compact support that vanish at $x=a,b$ along
with all their derivatives~\cite{ROBERTSCMP}:
\begin{eqnarray}
      \Czi :=
      \{ f \in  L^2(\mathbb R, \rmd x) \, : \ 
      f\in C^{\infty}(\mathbb R ) \, , f^{(n)}(a)=f^{(n)}(b)=0 \, ,
       \nonumber \\
      \hskip3.3cm  f \ \mbox{has compact support} \} \, .
\end{eqnarray}
Because
\begin{equation}
      \Czi \subset \Sw \, , 
\end{equation}
and because $\Czi$ is dense in $L^2(\mathbb R, \rmd x)$~\cite{ROBERTSCMP},
the space $\Sw$ is dense in $L^2(\mathbb R, \rmd x)$.

\vskip0.3cm

({\it iii}) From definition (\ref{definitionketE}), it is pretty easy 
to see that $|E ^{\pm}\rangle _{\rm l,r}$ are antilinear functionals. In order
to show that $|E ^{\pm}\rangle _{\rm l,r}$ are continuous, we define
\begin{equation}
      {\cal M}^{\pm}_{\rm l,r}(E):= \sup _{x\in \mathbb R} 
          \left| \chi ^{\pm}_{\rm l,r}(x;E) \right| 
\end{equation}
and
\begin{equation}
      C^{\pm}_{\rm l,r}(E):= {\cal M}^{\pm}_{\rm l,r}(E)
       \left( \int_{-\infty}^{\infty}\rmd x
      \, \frac{1}{(1+x^2)^2} \right) ^{1/2} \, .
\end{equation}
Since
\begin{eqnarray}
      |\langle \varphi |E^{\pm}\rangle _{\rm l,r}| &=&
       \left| \int_{-\infty}^{\infty}\rmd x \, 
         \overline{\varphi (x)}\chi^{\pm}_{\rm l,r}(x;E)\right| \nonumber \\
     &\leq & {\cal M}^{\pm}_{\rm l,r}(E) \int _{-\infty}^{\infty}
         \rmd x \, |\varphi (x)|
           \nonumber \\
      &=& {\cal M}^{\pm}_{\rm l,r}(E) \int_{-\infty}^{\infty}\rmd x \,
      \frac{1}{1+x^2} (1+x^2) |\varphi (x)| \nonumber \\
      &\leq & {\cal M}^{\pm}_{\rm l,r}(E) \left( \int_{-\infty}^{\infty}\rmd x
      \, \frac{1}{(1+x^2)^2} \right) ^{1/2} 
      \left( \int_{-\infty}^{\infty}\rmd x \, 
      \left| (1+x^2) \varphi (x) \right| ^2 \right) ^{1/2} \nonumber \\
      &=&C^{\pm}_{\rm l,r}(E) \,
          \| (1+Q^2)\varphi \| \nonumber \\ 
      &\leq &C^{\pm}_{\rm l,r}(E) \,
          (\| \varphi \| +  \|Q^2\varphi \|) \nonumber \\
      &=&C^{\pm}_{\rm l,r}(E) \,
         ( \| \varphi \|_{0,0,0} +  \|\varphi \| _{0,2,0}) \, ,
      \label{contekets}
\end{eqnarray}
the functionals $|E^{\pm}\rangle _{\rm l,r}$ are continuous when 
$\mathbf \Phi$ is endowed with the topology $\tau _{\mathbf \Phi}$. The
proof that $|p\rangle$ and $|x\rangle$ are also continuous antilinear 
functionals over $\mathbf \Phi$ is similar. 

\vskip0.3cm

({\it iv}) In order to prove that $|E^{\pm}\rangle _{\rm l,r}$ are 
generalized eigenvectors of $H$, we make use of the conditions 
(\ref{ddomain}) and (\ref{nmnorms}) satisfied by the elements of 
$\mathbf \Phi$,
\begin{eqnarray}
       \langle \varphi |H^{\times}|E^{\pm}\rangle _{\rm l,r} &=& 
      \langle H^{\dagger}\varphi |E^{\pm}\rangle _{\rm l,r}
       \nonumber \\
       &=& \int_{-\infty}^{\infty}\rmd x \, 
       \left( -\frac{\hbar ^2}{2m}\frac{\rmd ^2}{\rmd x^2}+V(x) \right)
       \overline{\varphi (x)} \chi ^{\pm}_{\rm l,r}(x;E) \nonumber \\
       &=&-\frac{\hbar ^2}{2m}
       \left[ \frac{\rmd \overline{\varphi (x)}}{\rmd x} 
          \chi ^{\pm}_{\rm l,r}(x;E)
       \right] _{-\infty}^{\infty} 
       +\frac{\hbar ^2}{2m}
       \left[ \overline{\varphi (x)} 
       \frac{\rmd \chi^{\pm}_{\rm l,r}(x;E)}{\rmd x} 
       \right] _{-\infty}^{\infty} \nonumber \\ 
       &&+ \int_{-\infty}^{\infty}\rmd x \, \overline{\varphi (x)}
       \left( -\frac{\hbar ^2}{2m}\frac{\rmd ^2}{\rmd x^2}+V(x) \right)
           \chi ^{\pm}_{\rm l,r}(x;E)
        \nonumber \\
      &=&E \int_{-\infty}^{\infty}\rmd x \, \overline{\varphi (x)}
        \chi ^{\pm}_{\rm l,r}(x;E)
        \nonumber \\
       &=&E\langle \varphi |E^{\pm}\rangle _{\rm l,r} \, .
         \label{ketsareegofH}
\end{eqnarray}

The proof that $|p\rangle$ and $|x\rangle$ are generalized eigenvectors of
$P$ and $Q$, respectively, is similar.

\end{proof}

\vskip0.5cm

\begin{proof}[Proof of Proposition~2] ({\it i}) It is clear from 
definitions (\ref{definitionbrap}), (\ref{definitionbrax}) and 
(\ref{definitionbraE}) that $\langle p|$, $\langle x|$ and 
$_{\rm l,r} \langle ^{\pm}E|$ are {\it linear} functionals over 
$\mathbf \Phi$. Because
\begin{eqnarray}
      \left| _{\rm l,r} \langle ^{\pm}E|\varphi \rangle \right| &=&
      \left| \langle \varphi |E^{\pm}\rangle _{\rm l,r} \right|  
       \hskip1cm \mbox{from (\ref{braketccE})}   \nonumber \\
      &\leq & C^{\pm}_{\rm l,r}(E) \,
         \left( \| \varphi \|_{0,0,0} +  \|\varphi \| _{0,2,0} \right) \, ,
        \hskip1cm \mbox{from (\ref{contekets})}
\end{eqnarray}
the bras $_{\rm l,r} \langle ^{\pm}E|$ are continuous. That $|p\rangle$ and 
$|x\rangle$ are also continuous over $\mathbf \Phi$ can be proved in a 
similar way. 

\vskip0.3cm

({\it ii}) Because
\begin{eqnarray}
     _{\rm l,r} \langle ^{\pm}E|H^{\prime}|\varphi \rangle &=&
     _{\rm l,r} \langle ^{\pm}E|H^{\dagger}\varphi \rangle \nonumber \\
       &=& \overline{\langle H^{\dagger}\varphi| E^{\pm}\rangle}_{\rm l,r} 
         \hskip1cm \mbox{from (\ref{braketccE})}    \nonumber \\
      &=& E \, \overline{\langle \varphi| E^{\pm}\rangle}_{\rm l,r}
           \hskip1cm  \mbox{from (\ref{ketsareegofH})}        \nonumber \\
      &=& E \hskip0.12cm _{\rm l,r} \langle ^{\pm}E|\varphi \rangle  \, ,
          \hskip1cm \mbox{from (\ref{braketccE})}
\end{eqnarray}
the bras $_{\rm l,r} \langle ^{\pm}E|$ are generalized left-eigenvectors of 
$H$.

Similarly, it can be proved that $\langle p|$ and $\langle x|$ are generalized 
left-eigenvectors of respectively $P$ and $Q$.
\end{proof}

\vskip0.5cm

\begin{proof}[Proof of Proposition~3] We only need to prove 
Eqs.~(\ref{spinofEbk})-(\ref{GMT2Q}), since 
Eqs.~(\ref{eigePket})-(\ref{eigeHket}) were proved in Proposition~1. Let us 
start with Eq.~(\ref{spinofEbk}). Take $\varphi$ and $\psi$ in 
$\Sw$. Because $U_{\pm}$ of 
Eq.~(\ref{dirintdec}) are unitary, we have that
\begin{equation}
       (\varphi ,\psi )=(U_{\pm}\varphi ,U_{\pm}\psi )=
       (\widehat{\varphi}^{\pm} ,\widehat{\psi}^{\pm} )   \, .
       \label{Usiuni}
\end{equation}
Since $\widehat{\varphi}^{\pm}$ and $\widehat{\psi}^{\pm}$ 
are in particular elements of 
$L^2([0,\infty ),\rmd E)\oplus L^2([0,\infty ),\rmd E)$, their scalar 
product is given by
\begin{equation}
      (\widehat{\varphi}^{\pm} ,\widehat{\psi}^{\pm})=
      \int_0^{\infty}\rmd E \, 
      \overline{\widehat{\varphi}_{\rm l}^{\pm}(E)}\, 
      \widehat{\psi}_{\rm l}^{\pm}(E) +
      \int_0^{\infty}\rmd E \, 
      \overline{\widehat{\varphi}_{\rm r}^{\pm}(E)}\, 
      \widehat{\psi}_{\rm r}^{\pm}(E) \, .
      \label{sphatvhaps}
\end{equation}
Since $\varphi$ and $\psi$ belong to $\Sw$, the 
actions of the eigenkets and eigenbras of $H$ are well defined on them:
\begin{eqnarray}
      \langle \varphi |E^{\pm} \rangle _{\rm l,r} =
            \overline{ \widehat{\varphi} _{\rm l,r}^{\pm}(E)} \, , 
       \label{actionofEphi}\\
   _{\rm l,r}\langle ^{\pm}E|\psi \rangle =
       \widehat{\psi} _{\rm l,r}^{\pm}(E) \, . \label{actionofEpsi}
\end{eqnarray}
By plugging Eqs.~(\ref{actionofEphi}) and (\ref{actionofEpsi}) into 
Eq.~(\ref{sphatvhaps}), and Eq.~(\ref{sphatvhaps}) into Eq.~(\ref{Usiuni}), 
we get to Eq.~(\ref{spinofEbk}). 

It is clear that the trick to prove (\ref{spinofEbk}) was to go to the
energy representation by way of $U_{\pm}$, in which representation $H$ acts 
as the multiplication operator. The same trick can be used to prove
Eq.~(\ref{GMT2H}). A similar trick applies to the proof of 
equations~(\ref{spinofpbk}) and (\ref{GMT2P}), although instead of $U_{\pm}$
we must use the Fourier transform to go to the momentum representation, where
$P$ acts as the multiplication operator. The calculations are 
straightforward and will not be reproduced here. Finally, 
Eqs.~(\ref{spinofxbk}) and (\ref{GMT2Q}) are immediate.
\end{proof}

\end{document}